\documentclass[journal,twoside,web]{ieeecolor}
\usepackage{tmi}
\usepackage{cite}
\usepackage{amsmath,amssymb,amsfonts}
\usepackage{algorithmic}
\usepackage{graphicx}
\usepackage{textcomp}
\usepackage{booktabs}
\usepackage{multirow}
\usepackage{rotating}
\usepackage[colorlinks,linkcolor=blue]{hyperref}
\usepackage[table]{xcolor}
\def\BibTeX{{\rm B\kern-.05em{\sc i\kern-.025em b}\kern-.08em
    T\kern-.1667em\lower.7ex\hbox{E}\kern-.125emX}}
\begin{document}
\title{Cohort-Individual Cooperative Learning for Multimodal Cancer Survival Analysis}
\author{Huajun Zhou, \IEEEmembership{Member, IEEE}, Fengtao Zhou, and Hao Chen, \IEEEmembership{Senior Member, IEEE}
\thanks{This work was supported by the Hong Kong Innovation and Technology Fund (Project No. ITS/028/21FP and No. PRP/034/22FX), Shenzhen Science and Technology Innovation Committee Fund (Project No. SGDX20210823103201011), and Research Grants Council of the Hong Kong Special Administrative Region, China (Project No. R6003-22 and C4024-22GF). (Corresponding author: Hao Chen.)}
\thanks{Huajun Zhou and Fengtao Zhou are with the Department of Computer Science and Engineering, The Hong Kong University of Science and Technology, Hong Kong, China (e-mail: csehjzhou@ust.hk; fzhouaf@connect.ust.hk).}
\thanks{Hao Chen is with the Department of Computer Science and Engineering, Department of Chemical and Biological Engineering and Division of Life Science, Hong Kong University of Science and Technology, Hong Kong, China (e-mail: jhc@cse.ust.hk). 
}}

\maketitle

\begin{abstract}
Recently, we have witnessed impressive achievements in cancer survival analysis by integrating multimodal data, \textit{e.g.}, pathology images and genomic profiles. 
However, the heterogeneity and high dimensionality of these modalities pose significant challenges in extracting discriminative representations while maintaining good generalization.
In this paper, we propose a Cohort-individual Cooperative Learning (CCL) framework to advance cancer survival analysis by collaborating knowledge decomposition and cohort guidance. 
Specifically, first, we propose a Multimodal Knowledge Decomposition (MKD) module to explicitly decompose multimodal knowledge into four distinct components: redundancy, synergy, and uniqueness of the two modalities. 
Such a comprehensive decomposition can enlighten the models to perceive easily overlooked yet important information, facilitating an effective multimodal fusion.
Second, we propose a Cohort Guidance Modeling (CGM) to mitigate the risk of overfitting task-irrelevant information.
It can promote a more comprehensive and robust understanding of the underlying multimodal data while avoiding the pitfalls of overfitting and enhancing the generalization ability of the model.
By cooperating with the knowledge decomposition and cohort guidance methods, we develop a robust multimodal survival analysis model with enhanced discrimination and generalization abilities.
Extensive experimental results on five cancer datasets demonstrate the effectiveness of our model in integrating multimodal data for survival analysis. 
Our code is available at \href{https://github.com/moothes/CCL-survival}{https://github.com/moothes/CCL-survival}.
\end{abstract}


\begin{figure}[!t]
\centering
\includegraphics[width=0.47 \textwidth]{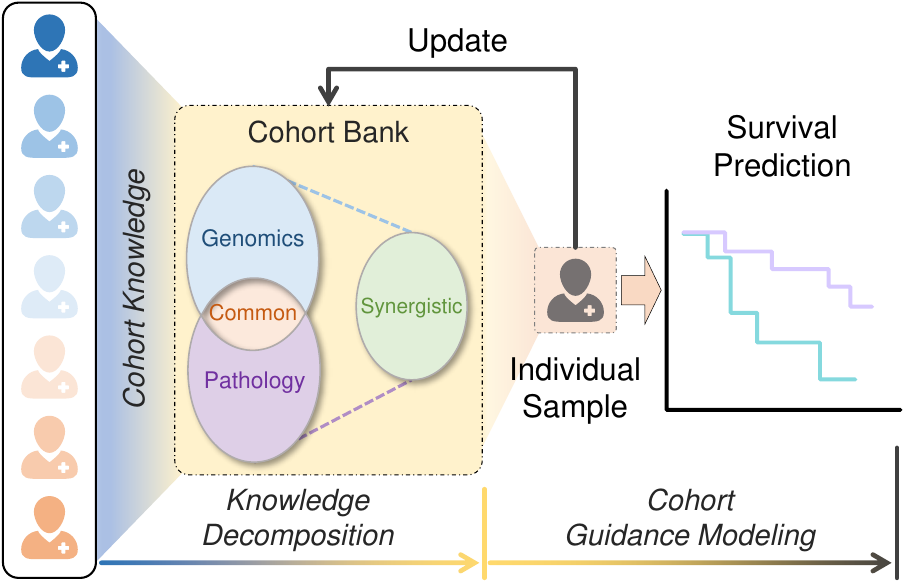}
\caption{Cohort knowledge offers a global view of multimodal data, assisting deep models to capture general multimodal interactions and facilitating a more effective fusion.}
\label{fig:motivation}
\end{figure}

\section{Introduction}
\label{sec:introduction}
\IEEEPARstart{S}{urvival} analysis, one of the most important tasks of cancer prognosis, aims to assess the probability of an event (typically death in survival analysis) occurring for a particular patient and accurately rank the risks of cancer patients.
It offers insights into disease progression, treatment effectiveness, and patient prognosis, ultimately leading to improved decision-making and patient care in research and clinical scenarios \cite{zhou2024multimodal}.
However, the complex nature of cancer necessitates a comprehensive evaluation of diverse personalized data, posing a significant challenge for survival analysis models to capture and incorporate this heterogeneity effectively.
Therefore, developing an effective multimodal integration approach is essential yet challenging for constructing robust and accurate survival analysis models.

Recent advances in Deep Learning (DL) \cite{alexnet, resnet, lin2024prompt} have made survival analysis more efficient and accurate by leveraging patients' clinical data \cite{surv2, surv1, surv5, surv6, surv3, surv4, geno1, geno2, geno3, geno4, path1, path2, path3, path4, klambauer2017self, wang2024rethinking}, \textit{e.g.}, genomics and pathology images, reducing clinicians' workloads considerably.
Genomic profiles provide molecular information, enabling personalized medicine and understanding of cancer genetics, while pathology images offer visual details that assist in diagnosis, grading, and assessing tumor heterogeneity. 
Together, these high-dimensional data enhance the understanding of tumors, thereby facilitating advancements in patient care and cancer research.
Recently, DL-based multimodal methods \cite{fam3l, hfbsurv, dof, pathfusion, xiong2024mome, smgt,gpdbn,rpgd} integrated these two modalities to enable the acquisition of complementary information from diverse perspectives, promising more accurate survival analysis.
For example, several works \cite{mcat, jaume2024modeling, chen2022pan, Xu_2023_ICCV, zhou2023cross, xu2024iclr} focused on enhancing modality representations by leveraging cross-modal interactions.
However, the challenges posed by the heterogeneity gap are still not well addressed, thus undermining the efficacy of multimodal integration.
Furthermore, the high dimensionality increases the risk of overfitting task-irrelevant information, resulting in performance degradation on unseen samples.

By tackling the above issues, our goal is to construct a more general and effective multimodal survival analysis model by incorporating a comprehensive knowledge decomposition and a more general patient cohort guidance.
First, the importance of different knowledge components in the integration process is variable.
For example, common knowledge shared by multiple modalities typically is redundant in multimodal integration, thus interfering with models to learn other discriminative information.
Furthermore, synergy is the new knowledge generated from multimodal interactions and may be overlooked without explicit modeling.
In our framework, we present a comprehensive decomposition of multimodal knowledge into four distinct components: redundancy, synergy, and uniqueness of the two modalities.
This decomposition will enable a deeper understanding of underlying factors impacting survival outcomes.
Second, extracting discriminative features from high-dimensional data while ensuring good generalization is a tough challenge.
For example, a wealth of task-irrelevant information can produce spurious correlations between modalities, demanding elaborated solutions to learn multimodal interactions with enhanced generalization ability.
Therefore, we seek to develop a more general patient cohort guidance mechanism, allowing for a broader range of patient characteristics to be considered during model training.
Such guidance can prevent the models from overemphasizing task-irrelevant information and thus enhance the model's generalization ability.
By combining these advancements, we can construct an effective and robust multimodal survival analysis model with enhanced accuracy and applicability.

In this paper, we propose a Cohort-individual Cooperative Learning (CCL) framework to integrate genomics and pathology images for cancer survival analysis.
Specifically, first, we propose a Multimodal Knowledge Decomposition (MKD) module to decompose multimodal knowledge into four distinct components: redundancy, synergy, and uniqueness of the two modalities. 
Such a comprehensive decomposition serves as an illuminating framework, enabling models to discern often disregarded yet crucial information. 
It paves the way for effective multimodal fusion, enhancing the integration of diverse data modalities and their complementary information.
Second, we propose a Cohort Guidance Modeling (CGM) to unleash the potential of distinct knowledge components and to enhance the model's generalization ability.
Our cohort guidance assists feature learning to at both knowledge and patient levels, capturing the essence of multifaceted data at various levels of granularity.
By cooperating with knowledge decomposition and cohort guidance, we enhance our model's discrimination and generalization abilities by effectively fusing diverse modalities while mitigating overfitting risks.
Experiment results on five datasets in The Cancer Genome Atlas (TCGA) program prove that our framework achieves state-of-the-art performance in survival analysis.

The main contributions are summarized as:
\begin{itemize}
    \item We propose a Multimodal Knowledge Decomposition (MKD) module to comprehensively and explicitly decompose multimodal knowledge into distinct components, facilitating an effective fusion of heterogeneous data. 
    \item We propose a Cohort Guidance Modeling (CGM) to enhance the generalization and discrimination abilities by mitigating the overfitting of task-irrelevant information.
    \item Extensive experiment results prove that the proposed framework achieves state-of-the-art performance on five datasets in The Cancer Genome Atlas (TCGA) program.
\end{itemize}

\section{Related Works}
\subsection{Unimodal Survival Analysis}
Survival analysis can be expressed as an estimation of the hazard function, which models the patient’s probability of death at a certain time, conditioned on personal clinical records.
In an early stage, Cox’s proportional hazards regression model \cite{cox} conceptualizes the hazard function as a multiplication of two components: 1) the underlying baseline hazard function, describing how the risk of event per time unit changes over time at baseline levels of covariates; and 2) the hazard ratio, measuring the impact of covariates.
Since the baseline hazard function for a given cancer is constant, the hazard ratio determines whether a patient is at high risk.
Based on Cox's regression model, subsequent survival analysis methods \cite{surv2, surv1, surv5, surv3, surv6, surv4} aim to predict the personalized hazard ratio using quantitative data extracted from short-term clinical indicators or long-term follow-up reports.
For example, Kappen et al. \cite{surv6} summarized seventeen pretreatment characteristics, such as residual tumor size, age, and thrombocytes, to predict the treatment outcome using a neural network.
Ohno-Machado et al. \cite{surv4} predicted the survival of AIDS patients based on demographics, laboratory markers, and clinical findings, such as age, hemoglobin, and albumin.

\textbf{Genomic profiles} provide molecular information on tumors, which is important for cancer prognosis prediction.
For example, certain genetic mutations or variations can affect tumor growth, metastasis, and response to chemotherapy or targeted therapies. 
Extensive studies \cite{geno1, geno2, geno4, geno3} have been conducted to predict cancer prognosis and treatment response by leveraging genomics data. 
Recently, Yousefi et al. \cite{geno4} combined deep learning and Bayesian optimization methods to tackle high-dimensional cancer outcomes prediction tasks.
Qiu et al. \cite{geno3} built good predictive models with limited high-dimensional samples by using a meta-learning survival analysis framework.

\begin{figure*}[t]
\centering
\includegraphics[width=0.99 \textwidth]{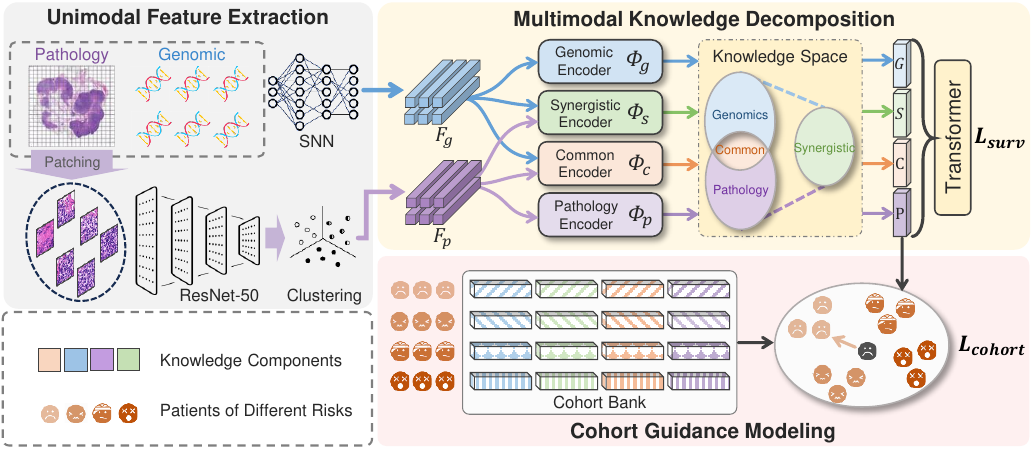}
\caption{An overview of the proposed Cohort-individual Cooperative Learning (CCL) strategy that learns comprehensive and discriminative knowledge components from multimodal data under the cohort guidance. Our CCL includes 1) a Multimodal Knowledge Decomposition (MKD) module to comprehensively and explicitly decompose heterogeneous multimodal knowledge; and 2) a Cohort Guidance Modeling (CGM) to learn more general and discriminative knowledge components.}
\label{fig:framework}
\end{figure*}

\textbf{Pathological images} provide morphological features of tumors, which can provide valuable information about the aggressiveness of the tumor, its response to treatment, and the likelihood of disease recurrence. 
Some recent works \cite{path1, path2, path3, path4} constructed effective survival analysis models based on giga-pixel Whole Slide Images (WSIs).
For example, Yao et al. \cite{path3} introduced the siamese MI-FCN and attention-based MIL pooling to efficiently learn imaging features from the WSI and then aggregate WSI-level information to the patient level.
Zhu et al. \cite{path4} adaptively sampled and grouped hundreds of patches from each WSI into several clusters. 
Then they employed an aggregation model to make patient-level predictions based on cluster-level survival prediction results.

Despite impressive performance achieved on survival analysis datasets, they have a narrow perspective and potentially overlook important aspects or correlations present in other modalities.
On the contrary, multimodal models can offer improved performance, robustness, and a more comprehensive understanding of cancer prognosis.

\subsection{Multimodal Survival Analysis}
Integrating genomics and pathology images can enhance the predictive power of survival analysis models \cite{fam3l, hfbsurv, dof, tfnet, smgt, pathfusion, rpgd, gpdbn, wang2024histo} and thus attract increasing attention recently.
By utilizing the cellular graph embedded in the tissue, Nakhli et al. \cite{smgt} extracted a unified representation for each patient, leveraging the hierarchical organization of the tissue.
Chen et al. \cite{pathfusion} captured the interactions between features across multiple modalities by combining the unimodal feature representations using the Kronecker product. 
Additionally, they incorporated a gating-based attention mechanism to control the expressive power of each representation.
Furthermore, Chen et al. \cite{mcat} proposed an interpretable, dense co-attention mapping between WSIs and genomic features formulated in the embedding space.
Xu et al. \cite{Xu_2023_ICCV} introduced the optimal transport theory to match WSI patches and gene embeddings for selecting informative patches to represent gigapixel WSIs and build an interpretable co-attention module to effectively fuse multimodal data.
Moreover, Zhou et al. \cite{zhou2023cross} found that the generated cross-modal representation can enhance and recalibrate intra-modal representation, and thus significantly improve its discrimination for survival analysis.

Existing co-attention-based methods focus on extracting common knowledge by using multimodal interactions.
Moreover, they are exposed to the risk of overfitting high-dimensional data, leading to performance degradation on unseen samples.
To address these issues, multimodal knowledge is comprehensively and explicitly decomposed in our framework to facilitate effective multimodal fusion.
Meanwhile, our framework leverages cohort guidance to improve the generalization ability of the decomposed knowledge components.

\section{Our Approach}
We propose a Cohort-individual Cooperative Learning (CCL) strategy to advance survival analysis by extracting more general and discriminative features, as depicted in Fig. \ref{fig:framework}. 
Our CCL includes a Multimodal Knowledge Decomposition (MKD) module to comprehensively and explicitly decompose multimodal knowledge into distinct components, and a Cohort Guidance Modeling (CGM) to enhance the generalization and discrimination abilities of our model.

\subsection{Unimodal Feature Extraction}
\textbf{Genomics.}
Genomic profiles can identify specific genetic alterations or biomarkers associated with cancer prognosis. 
For example, certain genetic mutations, gene expression patterns, or alterations in DNA copy number can serve as prognostic markers, helping to predict the likelihood of survival conditions of patients.
In our framework, we partition RNA sequencing (RNA-seq), Copy Number Variation (CNV), and Simple Nucleotide Variation (SNV) sequences into six sub-sequences as previous methods \cite{zhou2023cross,Xu_2023_ICCV}. 
Each sub-sequence $g_i$ is transformed into a feature by two cascaded 256-dimension Self-normalizing Neural Network (SNN) \cite{klambauer2017self} layers.
Finally, we obtain the genomic representation $F_{g} = fc_1(SNN(SNN(g_i)))$ by using a fully-connected layer $fc_1$ to aggregate the representations of multiple sequences into a single genomic representation $F_g \in \mathbb{R}^{1\times 256}$.

\begin{figure}[!t]
\centering
\includegraphics[width=0.47 \textwidth]{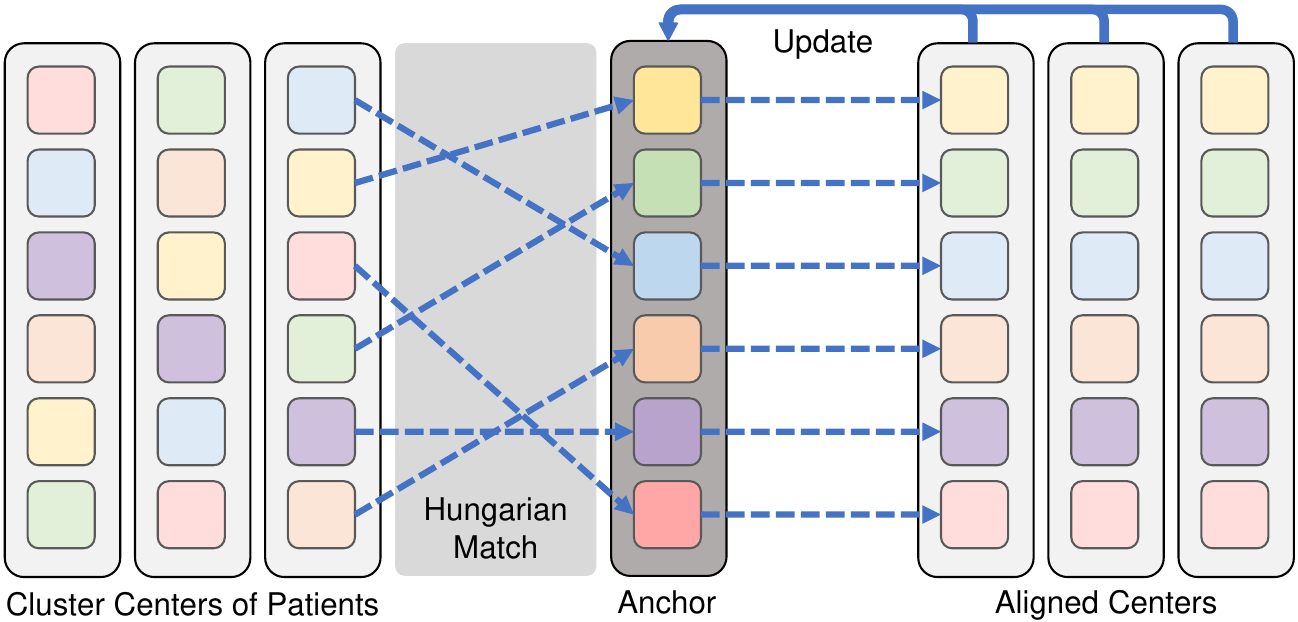}
\caption{Cluster center alignment for pathology features. Aligned centers at the same positions typically exhibit similar phenotypes across patients, enabling the network to extract specific information about these phenotypes.}
\label{fig:match}
\end{figure}

\textbf{Pathology.}
Whole Slide Images (WSIs), \textit{i.e.}, pathology images, describe the information about the tumor immune microenvironment and provide valuable information for cancer prognosis prediction. 
Considering the huge resolution of WSI, which exceeds the capacity of Convolutional Neural Networks (CNNs), we adopt a strategy of splitting tissue regions within each WSI into non-overlapping patches at 20x magnification (256 x 256 resolution for each patch).
Following previous works \cite{Xu_2023_ICCV,zhou2023cross,pathfusion}, we utilize an ImageNet pre-trained ResNet-50 \cite{resnet} model to extract a 1024-dimensional embedding for each patch, while all patch embeddings of the same WSI are collected as an embedding set.
It is worth noting that such an embedding set is still high-dimensional as it typically contains tens of thousands of patches for each WSI.
To further reduce information redundancy, we employ the K-means algorithm \cite{macqueen1967some} to cluster all patch embeddings into $k$ groups and leverage the cluster centers as pathology features.
However, due to the stochastic nature of K-means, cluster centers of different samples may be misaligned, which means that two cluster centers from the same ordinal position of two samples may exhibit completely different phenotypes. 
Consequently, deep models prefer to learn phenotype-independent knowledge instead of specialized knowledge tailored to each phenotype, overlooking crucial information in important patches. 
In our framework, we address this misalignment issue by cluster center alignment (CCA), an optimal matching between an anchor and cluster centers, as shown in Fig. \ref{fig:match}. 
This involves assigning each cluster center to a feature in the anchor with maximized similarities of matched pairs.
To solve this, the Hungarian algorithm \cite{kuhn1955hungarian} is employed to calculate the permutation matrix that maps the current centers to their matched ordinal positions. 
After that, we obtain the aligned centers by multiplying cluster centers with the permutation matrix.
The aligned centers are utilized to update the anchor with a ratio of $\tau$ during training.
Finally, a 256-dimension SNN layer and a fully-connected layer $fc_1$ are employed to aggregate the cluster centers into a single pathology representation $F_p \in \mathbb{R}^{1\times 256}$.
Overall, the pathology representation is generated by $F_p = fc_1(SNN(Matching(Kmeans(ResNet(p_i)), B)))$, where $p_i$ is the pathology patch and $B$ is the anchor.

\subsection{Cohort-individual Cooperative Learning}
Genomic profiles and pathology images provide both unique and collaborative insights into tumors.
Multimodal methods that can effectively integrate these two modalities are promised to be more impressive in learning discriminative representations than unimodal ones.
To achieve this goal, decomposing knowledge into modular components enables scalability and flexibility in handling high-dimensional and heterogeneous multimodal data. 
Previous multimodal methods \cite{li2023decoupled,hazarika2020misa} simply decompose multimodal knowledge into common and specific components.
However, synergy-new knowledge that can only be generated through multimodal collaborations-is valuable yet overlooked in these methods.
To better utilize all knowledge components in multimodal data, we propose a multimodal knowledge decomposition module and cohort guidance modeling to eliminate redundancy and strengthen valuable information under the guidance of patient cohorts.

\textbf{Multimodal Knowledge Decomposition.}
In our framework, we decompose the knowledge within genomic profiles and pathology images into four distinct components, including redundancy, synergy, and uniqueness of two modalities.
Concretely, the large heterogeneity gap between pathology and genomics indicates a wealth of modality-specific knowledge.
Meanwhile, there still exists considerable common knowledge between them.
For example, particular genetic mutations or alterations can be observed at both the molecular level and the histopathological level simultaneously.
Furthermore, another essential component, synergy, can provide new insights inon tumors to enrich the knowledge of multimodal data, especially for heterogeneous modalities.
Emerging from multimodal interactions, it surpasses common knowledge by forming a unique realm of insights unattainable through any individual modality alone.
As an example, the diagnosis of glioma typically necessitates genetic markers (IDH mutation and 1p/19q co-deletion) to categorize gliomas into subtypes and histopathological images to determine the presence of microvascular proliferation and necrosis simultaneously.
Overall, a comprehensive and explicit decomposition enhances integration, representation, synergy capture, and flexibility. 
It enables more effective modeling and utilization of multimodal data, leading to improved performance, insights, and decision-making in cancer prognosis.

\begin{figure}[!t]
\centering
\includegraphics[width=0.47 \textwidth]{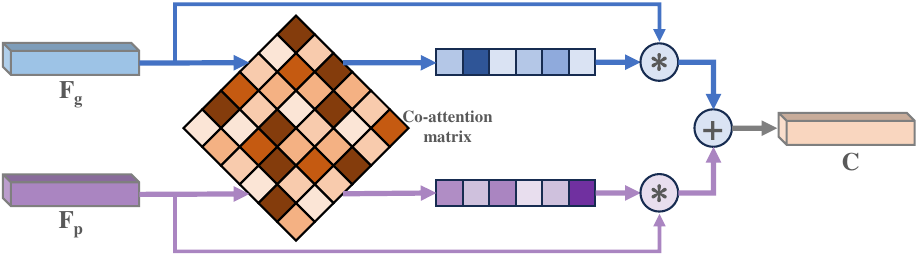}
\caption{The structure of common and synergistic encoders. We generate attention vectors based on the co-attention matrix to integrate multimodal features. The learnable parameters for common and synergistic knowledge are different, enabling the same structure used to extract distinct knowledge components. The $*$ and $+$ indicate element-wise multiplication and addition, respectively.}
\label{fig:coatt}
\end{figure}

To achieve a comprehensive knowledge decomposition, we develop four encoders within our framework. 
These encoders are specifically employed to capture and model different components—redundancy, synergy, and uniqueness~\cite{williams2010nonnegative, liang2023foundations, liang2024quantifying, liang2023multimodal}—present within the genomic profiles and pathology images.
Concretely, we employ MLP layers as modality encoders $\Phi_{p}$ and $\Phi_{g}$ to extract specific knowledge $P$ and $G$, focusing on aspects distinct from common knowledge shared by both modalities.
The formula can be written as:
\begin{equation}
    P = \Phi_{p}(F_{p}), G = \Phi_{g}(F_{g}).
\end{equation}
Diverging from $\Phi_{p}$ and $\Phi_{g}$ that receive only a single input, common and synergistic encoders $\Phi_{c}$ and $\Phi_{s}$ employed in this context necessitate two inputs to build multimodal interactions from different perspectives.
In these encoders, a co-attention block is utilized to bridge the interactions across modalities and produce modality attentions to integrate knowledge from different modalities into a fused feature $C$ or $S$, as shown in Fig \ref{fig:coatt}. 
The detailed operations can be formulated as:
\begin{align}
\centering
    C = \Phi_{c}(F_{p}, F_{g}) &= fc_1(A^T) * F_{p} + fc_1(A) * F_{g}, \\
    A &= fc(F_{p})^T\cdot fc(F_{g}),
\end{align}
where $A$ is the co-attention matrix, while $fc$ is a 256-dimension fully-connected layer to further fuse input features.
$fc_1$ is a fully-connected layer that transforms $256\times 256$ co-attention matrix into a $1\times 256$ attention vector for pathology and genomics representations, respectively.
The $\cdot$ means matrix multiplication, while $*$ indicates element-wise multiplication.
Synergy $S$ is computed the same as common knowledge $C$, but with different parameters.
Using the above encoders, we decompose modality representations $F_{g}$ and $F_{p}$ into genomic-specific $G$, pathology-specific $P$, common $C$, and synergistic $S$ components, respectively.

\begin{figure}[!t]
\centering
\includegraphics[width=0.47 \textwidth]{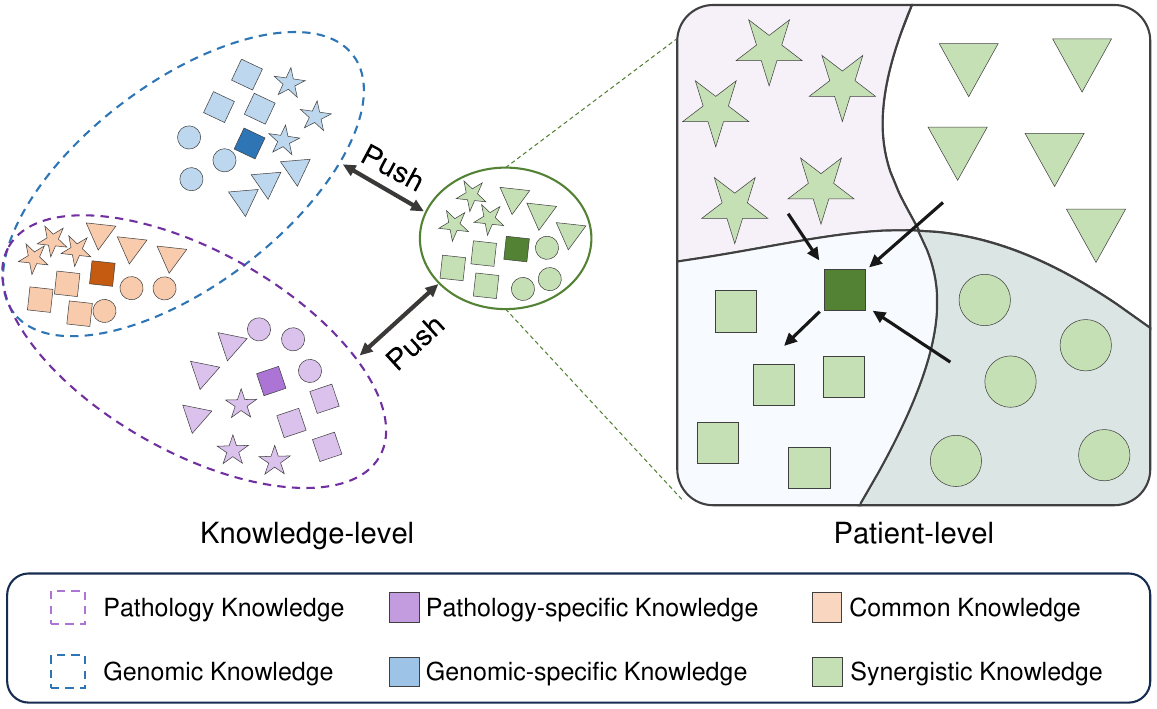}
\caption{Graphical illustration of cohort guidance. Colors and shapes indicate knowledge components and patient groups, respectively.}
\label{fig:cohort}
\end{figure}

\textbf{Cohort Guidance Modeling.}
In the above section, we utilize various encoders to capture diverse knowledge components from multimodal data. 
Nonetheless, relying solely on different modules does not guarantee the effective extraction of the intended knowledge components. 
To overcome this challenge, the incorporation of additional supervision signals becomes necessary. 
These signals serve to guide multiple encoders in acquiring distinct knowledge components, thereby facilitating the process of multimodal fusion.
Moreover, due to the high dimensionality of input modalities, multimodal interactions learned from each patient, as done in existing methods \cite{mcat,Xu_2023_ICCV,zhou2023cross}, may overfit task-irrelevant information, leading to reduced generalization and discrimination abilities.
Based on the above analysis, we propose to utilize the correlations between patient cohorts to acquire more general knowledge components with cohort consistency, capture the heterogeneity of multimodal data, and gain a better understanding of cancer.
In our framework, we harness the cohort guidance to learn knowledge components from both knowledge and patient levels, as shown in Fig. \ref{fig:cohort}.

\textit{At the knowledge level}, the most prominent difference between the decomposed knowledge components is their correlation to the original modality representations.
Specifically, synergy is unattainable through any individual modality alone, so it is beyond the knowledge of both modalities.
Redundancy is the intersection of knowledge as it is shared by both modalities.
In addition, uniqueness is located in the source modality and is distinct from another modality.
Therefore, we employ a set of similarity constraints between them:
\begin{equation}
    \begin{split}
    l_{k} =& |cos(G, F_p)| - cos(G, F_g) - cos(P, F_p)   \\ 
    +& |cos(P, F_g)| - cos(C, F_p) - cos(C, F_g) \\
    +& |cos(S, F_p)| + |cos(S, F_g)| ,
    \end{split}
\end{equation}
where $cos$ computes the cosine similarity between two inputs and $|\cdot|$ is the absolute operation used to constrain two inputs orthogonal to each other.
By minimizing $l_k$, multiple encoders are guided to learn distinct knowledge components.
In this context, the genomic-specific component $G$ captures knowledge specifically derived from genomics data, distinct from the knowledge obtained from pathology data. To achieve this, we maximize the similarity between the genomic-specific component $G$ and the genomics representation $F_g$ by using the term $-cos(G, F_g)$. Simultaneously, we enforce orthogonality between $G$ and the pathology representation $F_p$ by using the term $|cos(G, F_p)|$.
Similarly, the pathology-specific component $P$ follows a similar guidance strategy as $G$ to capture knowledge specific to pathology data.
Furthermore, there exists common knowledge shared by both modalities. To model this shared knowledge, we maximize its similarities to both the genomics representation $F_g$ and the pathology representation $F_p$ by using the term $-cos(C, F_p) - cos(C, F_g)$.
In addition, the synergistic component $S$ is supervised to minimize its similarities to both the genomics and pathology representations. Since it represents knowledge beyond the scope of both modalities, we use the term $|cos(S, F_p)| + |cos(S, F_g)|$ to achieve this objective.
We freeze the gradients of $F_p$ and $F_g$ from $l_k$ to prevent the potential model collapse that the extracted features follow similar distributions.

\textit{At the patient level}, task-relevant information can be obtained by distinguishing patients conditioned on their risk scores, which are valuable for extracting more discriminative representations.
Therefore, we split all patients into $r$ equal groups conditioned on their ground truth survival time, while patients in the same group are considered as similar.
For uncensored patients, its features are closer to the features extracted from similar patients, while distinctive to patients in other groups.
For censored patients, it may be closer to the features of patients at similar or lower risk and distinctly different from those at higher risk.
Such an assumption is in line with the definition of contrastive learning, which has been proven to have strong feature learning capability in both supervised and unsupervised settings.
Therefore, we develop a cohort contrastive learning to enhance the discrimination and generalization abilities of our model.
In the training phase, we build a cohort bank to store features of historical patients in each group.
The cohort bank is implemented as a queue, following the ``First In, First Out'' (FIFO) principle.
Specifically, given the bank length $b$ and current iteration $t$, the cohort bank stores the decomposed components from iteration ($t-b$) to ($t-1$).
After the current iteration, the cohort bank will be updated to store the decomposed components from iteration ($t-b+1$) to ($t$).
It is noteworthy that when $t<b+1$, no representations in the cohort bank will be discarded when updating.

For the feature of uncensored patients, we reduce its distances to similar patients in the cohort bank, while enlarging its distances to other patients.
As an example, the formula for synergy $S$ is:
\begin{equation}
    l_{p} = - \log \frac{\sum_{S' \in \mathcal{S}_{+}} d(S, S')}{\sum_{S' \in \mathcal{S}_{+}} d(S, S') + \sum_{S' \in \mathcal{S}_{-}} d(S, S')},
\end{equation}
where $\mathcal{S}_+$ is the patient group of similar risk storing in the cohort bank, while $\mathcal{S}_-$ is the collection of other patients.
$d$ is the similarity function.
Note that $S$ can be replaced by other knowledge components $G$, $P$, or $C$ similarly.
For censored patients, we can extend $\mathcal{S}_+$ as the collection of patients with similar or lower risk.

Combining knowledge and patient levels, the cohort guidance loss can be formulated as:
\begin{equation}
    L_{cohort} = l_k + l_p.
\end{equation}

\subsection{Multimodal Fusion and Prediction}
Recently, Transformer \cite{transformer} has shown impressive performance on multimodal learning, and thus is employed to integrate the decomposed components for survival predictions.
Specifically, we concatenate a class token $U$ with the decomposed components $G$, $P$, $C$, and $S$ as the inputs of Transformer $\Phi_t$.
Moreover, a fully-connected layer with Sigmoid activation, denoted as $\sigma$, is appended to the output of the class token for predicting the hazard function $H$:
\begin{equation}
    H=\sigma(\Phi_t(U, G, P, S, C)).
\end{equation}

For survival prediction, following previous works \cite{mcat,jaume2024modeling,Xu_2023_ICCV,zhou2023cross}, we construct a one-hot time-series vector $T$ based on $n$ equal parts of ground truth survival time.
Under this setting, the original event time regression problem is simplified to a classification problem. 
We find out the time interval $t_k$ in which the ground truth event occurred as the classification label $k$ for each patient.
Each patient sample is defined as a triplet $\{H, c, k\}$, where $H=\{h_1,...h_n\}$ is the predicted hazard vector measuring the probability of ground truth event time located in corresponding time interval $t$.
Additionally, we define the discrete survival function $f_{sur}(H,k)=\prod_{j=1}^{k}(1-h_{j})$.
Following previous works \cite{Xu_2023_ICCV,zhou2023cross,mcat}, we generalize the negative log-likelihood (NLL) with censorship to supervise the survival prediction by:
\begin{equation}
    \begin{split}
    L_{surv}=&-c\log(f_{sur}(H,k))\\
    &-(1-c)\log(f_{sur}(H,{k-1}))\\
    &-(1-c)\log(h_k).
    \end{split}
\end{equation}
Finally, the overall loss function of our framework is:
\begin{equation}
    L=L_{surv}+ \alpha L_{cohort},
\end{equation}
where $\alpha$ controls the effect of our cohort guidance.

\begin{table*}[ht]
\centering
\caption{Survival analysis (C-index) on TCGA datasets. The best and second best scores are in \textbf{bold} and \underline{underline}, respectively. All results are re-produced on our own environment with the same settings.}
\setlength\tabcolsep{9pt}
   \begin{center}
      \begin{tabular}{l|c|ccccc|c}
        \toprule
        \multirow{2}{*}{Methods} &\multirow{2}{*}{Modality} & \multicolumn{5}{c|}{Datasets} &\multirow{2}{*}{Overall}\\\cline{3-7}
        &   & BLCA   & BRCA  & GBMLGG & LUAD  & UCEC       &      \\
         \midrule
        \multicolumn{8}{l}{\textbf{Pathology feature extraction backbone -- ResNet-50 \cite{resnet}}} \\\hline
         SNN~\cite{klambauer2017self}      & Genomics                                            & 0.6339$_{\pm0.0509}$               & 0.6327$_{\pm0.0739}$               & 0.8370$_{\pm0.0276}$               & 0.6171$_{\pm0.0411}$               & 0.6900$_{\pm0.0389}$          & 0.6821      \\
         SNNTrans & Genomics                                           & 0.6456$_{\pm0.0428}$               & 0.6478$_{\pm0.0580}$               & 0.8284$_{\pm0.0158}$               & 0.6335$_{\pm0.0493}$               & 0.6324$_{\pm0.0324}$          & 0.6775       \\
         \midrule
         MaxMIL                            & Pathology                                  & 0.5509$_{\pm0.0315}$               & 0.5966$_{\pm0.0547}$               & 0.7136$_{\pm0.0574}$               & 0.5958$_{\pm0.0600}$               & 0.5626$_{\pm0.0547}$           & 0.6039     \\
         MeanMIL                           & Pathology                                  & 0.5847$_{\pm0.0324}$               & 0.6110$_{\pm0.0286}$               & 0.7896$_{\pm0.0367}$               & 0.5763$_{\pm0.0536}$               & 0.6653$_{\pm0.0457}$           & 0.6454     \\
         AttMIL~\cite{ilse2018attention}   & Pathology                                   & 0.5673$_{\pm0.0498}$               & 0.5899$_{\pm0.0472}$               & 0.7974$_{\pm0.0336}$               & 0.5753$_{\pm0.0744}$               & 0.6507$_{\pm0.0330}$             & 0.6361   \\
         CLAM-SB~\cite{lu2021data}         & Pathology                                   & 0.5487$_{\pm0.0286}$               & 0.6091$_{\pm0.0329}$               & 0.7969$_{\pm0.0346}$               & 0.5962$_{\pm0.0558}$               & 0.6780$_{\pm0.0342}$       & 0.6458         \\
         CLAM-MB~\cite{lu2021data}         & Pathology                                  & 0.5620$_{\pm0.0313}$               & 0.6203$_{\pm0.0520}$               & 0.7986$_{\pm0.0320}$               & 0.5918$_{\pm0.0591}$               & 0.6821$_{\pm0.0646}$        & 0.6510        \\
         TransMIL~\cite{shao2021transmil}  & Pathology                                  & 0.5466$_{\pm0.0334}$               & 0.6430$_{\pm0.0368}$               & 0.7916$_{\pm0.0272}$               & 0.5788$_{\pm0.0303}$               & 0.6799$_{\pm0.0304}$       & 0.6480          \\
         DTFD~\cite{dtfd}                  & Pathology                                  & 0.5662$_{\pm0.0353}$               & 0.5975$_{\pm0.0406}$               & 0.7641$_{\pm0.0297}$               & 0.5580$_{\pm0.0404}$               & 0.6308$_{\pm0.0190}$       & 0.6233         \\
         \midrule
         DualTrans                         & Multimodal                                & 0.6607$_{\pm0.0319}$               & 0.6637$_{\pm0.0621}$ & 0.8393$_{\pm0.0174}$               & 0.6706$_{\pm0.0343}$ & 0.6724$_{\pm0.0192}$         & 0.7013       \\
         MCAT~\cite{mcat}    & Multimodal                                & 0.6727$_{\pm0.0320}$ & 0.6590$_{\pm0.0418}$               & 0.8350$_{\pm0.0233}$               & 0.6597$_{\pm0.0279}$               & 0.6336$_{\pm0.0506}$          & 0.6920      \\
         M3IF~\cite{li2021multi}           & Multimodal                              & 0.6361$_{\pm0.0197}$               & 0.6197$_{\pm0.0707}$               & 0.8238$_{\pm0.0170}$               & 0.6299$_{\pm0.0312}$               & 0.6672$_{\pm0.0293}$      & 0.6753          \\
         GPDBN~\cite{gpdbn}        & Multimodal                                 & 0.6354$_{\pm0.0252}$               & 0.6549$_{\pm0.0332}$        &        0.8510$_{\pm0.0243}$ & 0.6400$_{\pm0.0478}$               & 0.6839$_{\pm0.0529}$          & 0.6930      \\
         Porpoise~\cite{chen2022pan}       & Multimodal                                & 0.6461$_{\pm0.0338}$               & 0.6207$_{\pm0.0544}$               & 0.8479$_{\pm0.0128}$               & 0.6403$_{\pm0.0412}$               & 0.6918$_{\pm0.0488}$  & 0.6894 \\
         HFBSurv~\cite{hfbsurv}      & Multimodal                                & 0.6398$_{\pm0.0277}$               & 0.6473$_{\pm0.0346}$               & 0.8383$_{\pm0.0128}$               & 0.6501$_{\pm0.0495}$               & 0.6421$_{\pm0.0445}$          & 0.6835      \\
         SurvPath~\cite{jaume2024modeling}     & Multimodal                                & 0.6581$_{\pm0.0357}$               & 0.6306$_{\pm0.0340}$               & 0.8422$_{\pm0.0161}$               & 0.6600$_{\pm0.0233}$               & 0.6636$_{\pm0.0354}$         & 0.6909       \\
         MOTCat~\cite{Xu_2023_ICCV}     & Multimodal                                & 0.6830$_{\pm0.0260}$               & \underline{0.6730$_{\pm0.0060}$}               & 0.8490$_{\pm0.0280}$               & 0.6700$_{\pm0.0380}$               & 0.6750$_{\pm0.0400}$         & 0.7100       \\
         CMTA~\cite{zhou2023cross}     & Multimodal                                & \textbf{0.6910$_{\pm0.0426}$}               & 0.6679$_{\pm0.0434}$              & \underline{0.8531$_{\pm0.0116}$}               & \underline{0.6864$_{\pm0.0359}$}               & \underline{0.6975$_{\pm0.0409}$}         & \underline{0.7192}       \\
          Ours       & Multimodal                                & \underline{0.6862$_{\pm0.0253}$}  & \textbf{0.6840$_{\pm0.0339}$}  & \textbf{0.8614$_{\pm0.0149}$}  & \textbf{0.6957$_{\pm0.0231}$}  & \textbf{0.7026$_{\pm0.0475}$}   & \textbf{0.7260} \\

         \midrule
         \midrule
          \multicolumn{8}{l}{\textbf{Pathology feature extraction backbone -- HIPT \cite{chen2022scaling}}} \\\hline
          MCAT~\cite{mcat}    & Multimodal                                & 0.6483$_{\pm0.0218}$ & 0.6364$_{\pm0.0277}$               & 0.8559$_{\pm0.0089}$               & 0.6587$_{\pm0.0254}$               & 0.6715$_{\pm0.0602}$          & 0.6941      \\
          SurvPath~\cite{jaume2024modeling}     & Multimodal                                & 0.6295$_{\pm0.0295}$               & 0.5821$_{\pm0.0675}$               & 0.8474$_{\pm0.0098}$               & 0.6384$_{\pm0.0326}$               & 0.6457$_{\pm0.0589}$         & 0.6686       \\
          MOTCat~\cite{Xu_2023_ICCV}     & Multimodal                                & \underline{0.6556$_{\pm0.0292}$}               & \textbf{0.6714$_{\pm0.0446}$}               & 0.8514$_{\pm0.0107}$               & 0.6694$_{\pm0.0371}$               & \textbf{0.7044$_{\pm0.0643}$}         & \underline{0.7104}       \\
          CMTA~\cite{zhou2023cross}     & Multimodal                                & 0.6542$_{\pm0.0373}$               & 0.6545$_{\pm0.0784}$              & \underline{0.8657$_{\pm0.0149}$}              & \textbf{0.6872$_{\pm0.0129}$}             & 0.6787$_{\pm0.0449}$         & 0.7081       \\
          Ours       & Multimodal                                & \textbf{0.6651$_{\pm0.0251}$}  & \underline{0.6621$_{\pm0.0452}$}  & \textbf{0.8667$_{\pm0.0098}$}  & \underline{0.6775$_{\pm0.0226}$}  & \underline{0.6933$_{\pm0.0408}$}   & \textbf{0.7124} \\
          
         \midrule
         \midrule
          \multicolumn{8}{l}{\textbf{Pathology feature extraction backbone -- PLIP \cite{huang2023visual}}} \\\hline
          MCAT~\cite{mcat}    & Multimodal                                & 0.6549$_{\pm0.0175}$ & 0.6761$_{\pm0.0456}$               & 0.8388$_{\pm0.0240}$               & 0.6651$_{\pm0.0163}$               & 0.7029$_{\pm0.0519}$          & 0.7076      \\
          SurvPath~\cite{jaume2024modeling}     & Multimodal                                & 0.6382$_{\pm0.0483}$               & 0.6252$_{\pm0.0317}$               & 0.8438$_{\pm0.0208}$               & 0.6599$_{\pm0.0448}$               & 0.6938$_{\pm0.0747}$         & 0.6922       \\
          MOTCat~\cite{Xu_2023_ICCV}     & Multimodal                                & \underline{0.6628$_{\pm0.0354}$}               & 0.6698$_{\pm0.0664}$               & 0.8381$_{\pm0.0144}$               & \textbf{0.6796$_{\pm0.0381}$}               & 0.6845$_{\pm0.0480}$         & 0.7069       \\
          CMTA~\cite{zhou2023cross}     & Multimodal                                & 0.6586$_{\pm0.0192}$               & \textbf{0.6834$_{\pm0.0559}$}              & \underline{0.8515$_{\pm0.0271}$}               & 0.6701$_{\pm0.0241}$               & \underline{0.7300$_{\pm0.0526}$}         & \underline{0.7187}       \\
          Ours       & Multimodal                                & \textbf{0.6633$_{\pm0.0445}$}  & \underline{0.6797$_{\pm0.0695}$}  & \textbf{0.8578$_{\pm0.0235}$}  & \underline{0.6784$_{\pm0.0383}$}  & \textbf{0.7309$_{\pm0.0478}$}   & \textbf{0.7220} \\
          
         \midrule
         \midrule
          \multicolumn{8}{l}{\textbf{Pathology feature extraction backbone -- CONCH \cite{lu2024visual}}} \\\hline
          MCAT~\cite{mcat}    & Multimodal                                & 0.6605$_{\pm0.0159}$ & 0.6372$_{\pm0.0571}$               & 0.8438$_{\pm0.0117}$               & 0.6640$_{\pm0.0275}$               & 0.7075$_{\pm0.0373}$          & 0.7026      \\
          SurvPath~\cite{jaume2024modeling}     & Multimodal                                & 0.6422$_{\pm0.0194}$               & 0.6740$_{\pm0.0445}$               & 0.8447$_{\pm0.0237}$               & 0.6438$_{\pm0.0493}$               & \underline{0.7477$_{\pm0.0333}$}         & 0.7105       \\
          MOTCat~\cite{Xu_2023_ICCV}     & Multimodal                                & 0.6662$_{\pm0.0327}$               & 0.6526$_{\pm0.0294}$               & 0.8442$_{\pm0.0188}$               & 0.6736$_{\pm0.0371}$               & 0.7160$_{\pm0.0319}$         & 0.7105       \\
          CMTA~\cite{zhou2023cross}     & Multimodal                                & \underline{0.6694$_{\pm0.0148}$}               & \textbf{0.7273$_{\pm0.0613}$}              & \textbf{0.8518$_{\pm0.0206}$}               & \underline{0.6750$_{\pm0.0100}$}               & 0.7431$_{\pm0.0484}$         & \textbf{0.7333}       \\
          Ours       & Multimodal                                & \textbf{0.6701$_{\pm0.0323}$}  & \underline{0.6958$_{\pm0.0431}$}  & \underline{0.8454$_{\pm0.0186}$}  & \textbf{0.6868$_{\pm0.0253}$}  & \textbf{0.7649$_{\pm0.0343}$}   & \underline{0.7326} \\
          
         \midrule
         \midrule
          \multicolumn{8}{l}{\textbf{Pathology feature extraction backbone -- UNI \cite{chen2024uni}}} \\\hline
          MCAT~\cite{mcat}    & Multimodal                                & 0.6510$_{\pm0.0291}$ & 0.6640$_{\pm0.0425}$               & 0.8515$_{\pm0.0289}$               & 0.6566$_{\pm0.0306}$               & 0.7007$_{\pm0.0556}$          & 0.7047      \\
          SurvPath~\cite{jaume2024modeling}     & Multimodal                                & 0.6290$_{\pm0.0376}$               & 0.7005$_{\pm0.0521}$               & 0.8286$_{\pm0.0204}$               & 0.6321$_{\pm0.0550}$               & 0.7376$_{\pm0.0475}$         & 0.7056       \\
          MOTCat~\cite{Xu_2023_ICCV}     & Multimodal                                & \textbf{0.6525$_{\pm0.0280}$}               & 0.6651$_{\pm0.0421}$               & 0.8491$_{\pm0.0154}$               & \underline{0.6684$_{\pm0.0275}$}               & 0.7383$_{\pm0.0552}$         & 0.7147       \\
          CMTA~\cite{zhou2023cross}     & Multimodal                                & \underline{0.6521$_{\pm0.0319}$}               & \underline{0.7059$_{\pm0.0304}$}             & \textbf{0.8541$_{\pm0.0207}$}               & 0.6642$_{\pm0.0462}$               & \underline{0.7403$_{\pm0.0432}$}         & \underline{0.7233}       \\
           Ours       & Multimodal                                & 0.6404$_{\pm0.0314}$  & \textbf{0.7152$_{\pm0.0410}$}  & \underline{0.8536$_{\pm0.0126}$}  & \textbf{0.6737$_{\pm0.0395}$}  & \textbf{0.7613$_{\pm0.0482}$}   & \textbf{0.7288} \\
         \bottomrule
      \end{tabular}
   \end{center}
   \label{tab:main}
\end{table*}

\section{Experiments}
\subsection{Experiment Setups}
\textbf{Dataset.} 
To validate the effectiveness of the proposed methods, we evaluate our framework on five datasets from TCGA: Bladder Urothelial Carcinoma (BLCA, n=372), Breast Invasive Carcinoma (BRCA, n=956), Glioblastoma \& Lower Grade Glioma (GBMLGG, n=569), Lung Adenocarcinoma (LUAD, n=453) and Uterine Corpus Endometrial Carcinoma (UCEC, n=480).
These datasets contain hundreds of paired genomics, pathological images, and follow-up data collected from multiple centers, enabling center-agnostic analysis on precision oncology.
We collected all diagnostic WSIs used for primary diagnosis, resulting in 2,830 WSIs with an average of 15k patches per WSI at 20x magnification (assuming 256 x 256 patches). 
Genomic profiles provide molecular information about individuals, including RNA sequencing (RNA-seq), Copy Number Variation (CNV), Simple Nucleotide Variation (SNV), and DNA methylation.
Following previous works \cite{mcat,zhou2023cross}, we use RNA-seq, CNV, and SNV sequences and further group them into six sub-sequences: 1) Tumor Suppression, 2) Oncogenesis, 3) Protein Kinases, 4) Cellular Differentiation, 5) Transcription, and 6) Cytokines and Growth.

\textbf{Evaluation Metrics.}
The concordance index (C-index) is the proportion of all comparable patient pairs for which the predicted outcome is consistent with the actual outcome. 
In survival analysis, the predicted outcome is considered consistent with the actual outcome if the predicted survival time is longer for individuals who indeed have a longer survival time, compared to the predicted event time for others.
We also calculate the standard deviations of C-index scores across five folds.
Moreover, Kaplan-Meier and T-test analyses are employed to assess the significance of differences in survival predictions between high- and low-risk groups.

\textbf{Implementation.}
For each dataset, we adopted five-fold cross-validation to evaluate our model and other compared methods.
Specifically, first, we shuffle the dataset randomly and split the dataset into five groups.
For each group, we take this group as a test data set, while the remaining groups as a training data set.
We fit the models on the training set and evaluate them on the test set to report the C-index scores.
Our framework was implemented using PyTorch running on a single NVIDIA GTX 3090 GPU. 
For optimization, we employed the SGD optimizer with a learning rate of 1e-3 to train our framework for 30 epochs.
We unify the settings of compared models to make a fair comparison and report the re-produced results in our environment.
For the initialization of random seed, we set it to 0 in all experiments.

\subsection{Comparative Methods}
We compare our framework with different types of survival analysis models, including genomics-based (SNN \cite{klambauer2017self}), pathology-based (AttMIL \cite{ilse2018attention}, CLAM \cite{lu2021data}, TransMIL \cite{shao2021transmil}, and DTFD \cite{dtfd}), and multimodal (MCAT \cite{mcat}, M3IF \cite{li2021multi}, GPDBN \cite{gpdbn}, Porpoise \cite{chen2022pan}, HFBSurv \cite{hfbsurv}, SurvPath \cite{jaume2024modeling}, MOTCat \cite{Xu_2023_ICCV}, and CMTA \cite{zhou2023cross}) models. 
Here we provide brief overviews of several representative multimodal competitors:
\begin{itemize}
\item \textbf{MCAT} \cite{mcat} is a multimodal fusion model that utilizes the co-attention between genomic and pathology features as the attention vector to select informative patches in WSIs. After that, two Transformers are employed to fuse genomic and pathology features, respectively. Finally, two features are concatenated to produce survival predictions.
\item \textbf{SurvPath} \cite{jaume2024modeling} learns biological pathway tokens from transcriptomics that can encode specific cellular functions and fuses two modalities using a memory-efficient multimodal Transformer to model interactions between pathway and histology patch tokens.

\begin{figure*}[t]
\centering
\begin{minipage}{1 \textwidth}
    \begin{rotate}{90}
    \begin{minipage}{0.18 \textwidth} \centering Ours \end{minipage}
    \end{rotate}
    \includegraphics[width=0.195 \textwidth]{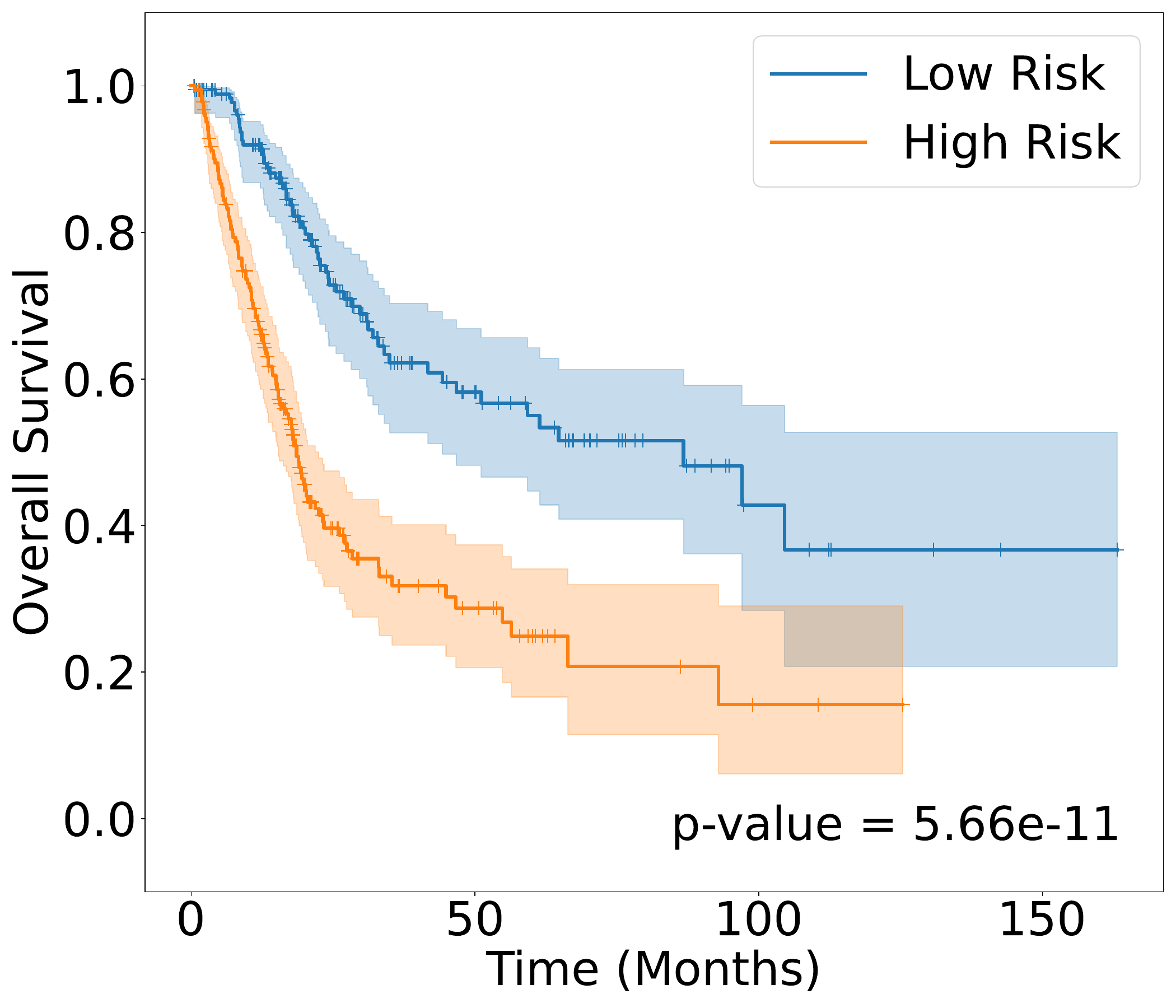}
    \includegraphics[width=0.195 \textwidth]{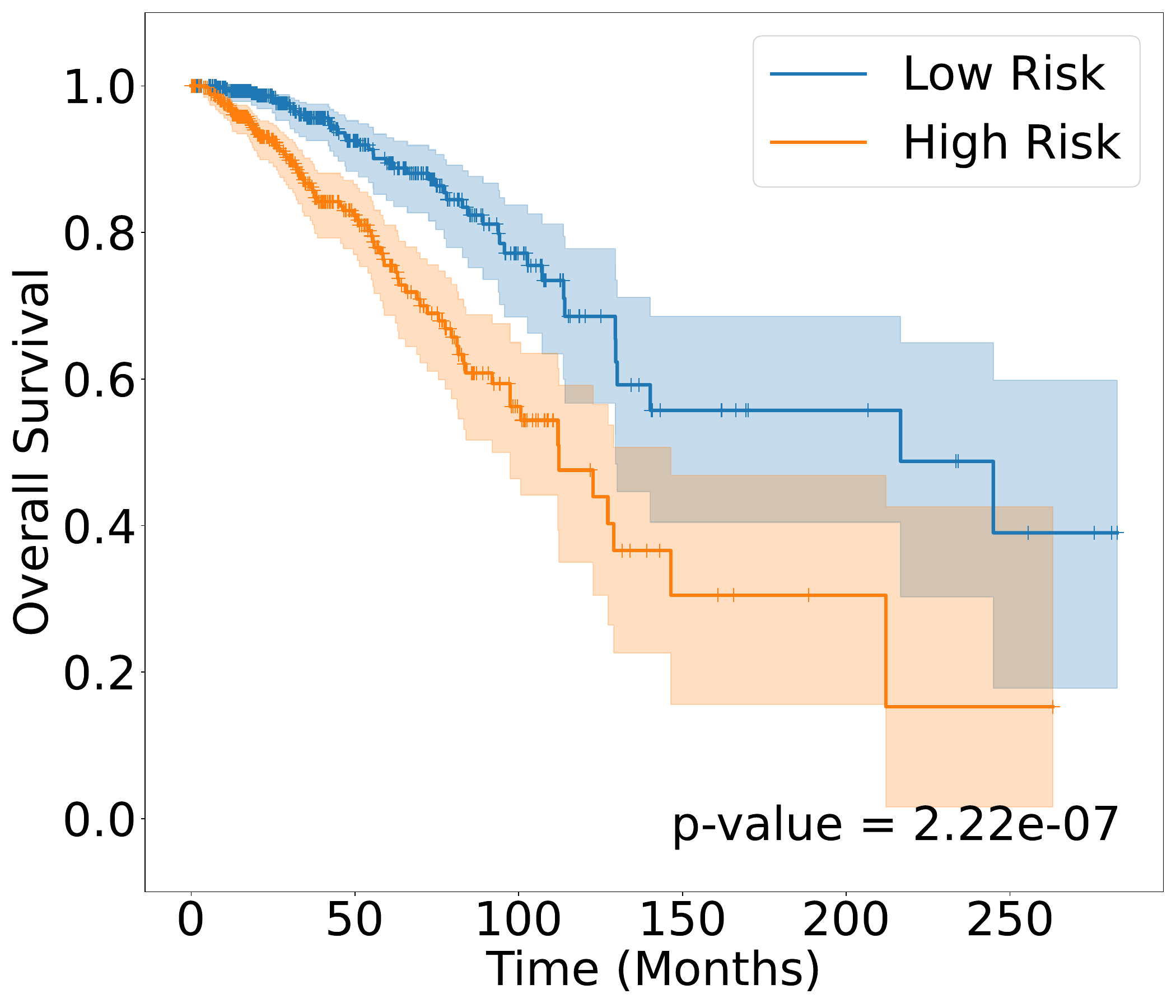}
    \includegraphics[width=0.195 \textwidth]{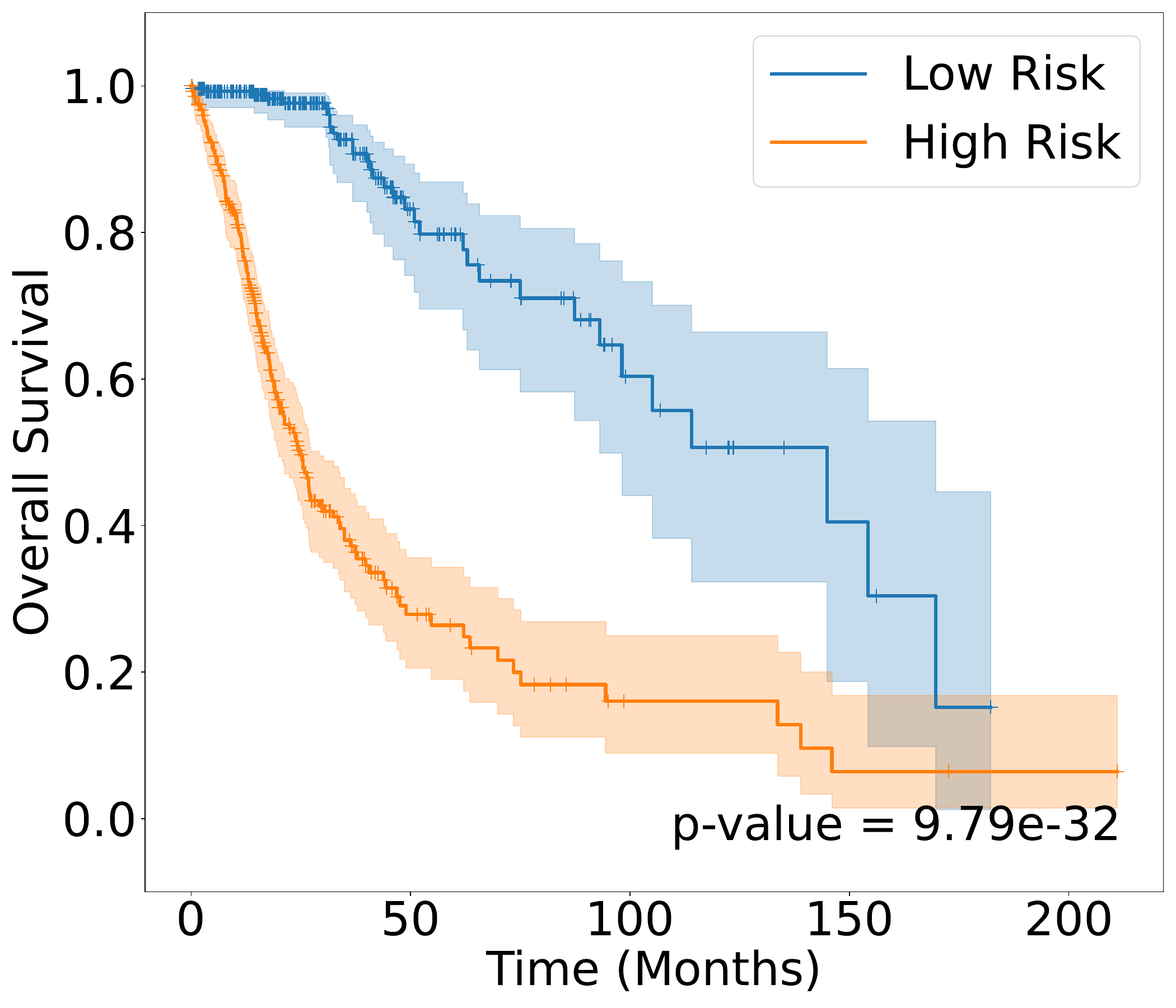}
    \includegraphics[width=0.195 \textwidth]{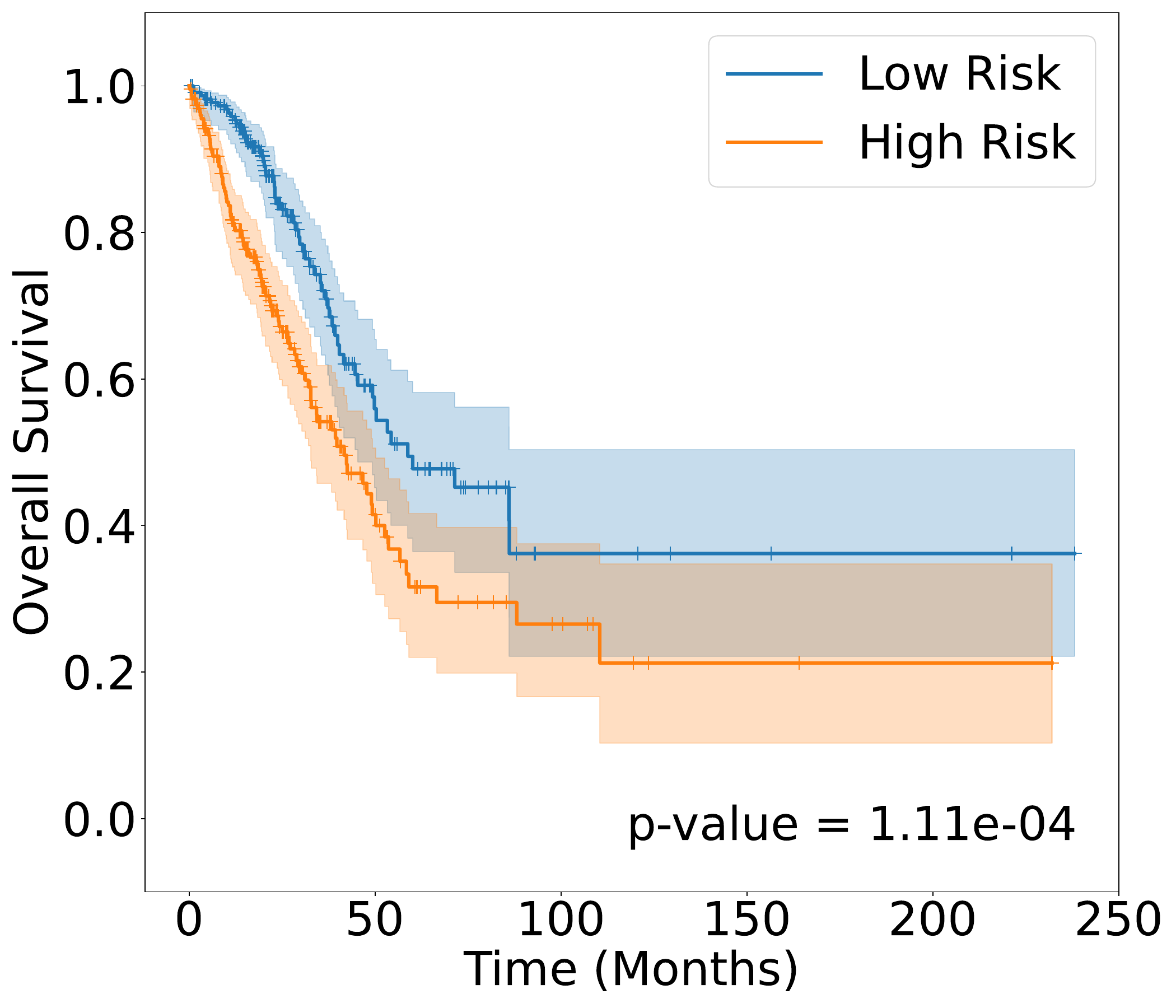}
    \includegraphics[width=0.195 \textwidth]{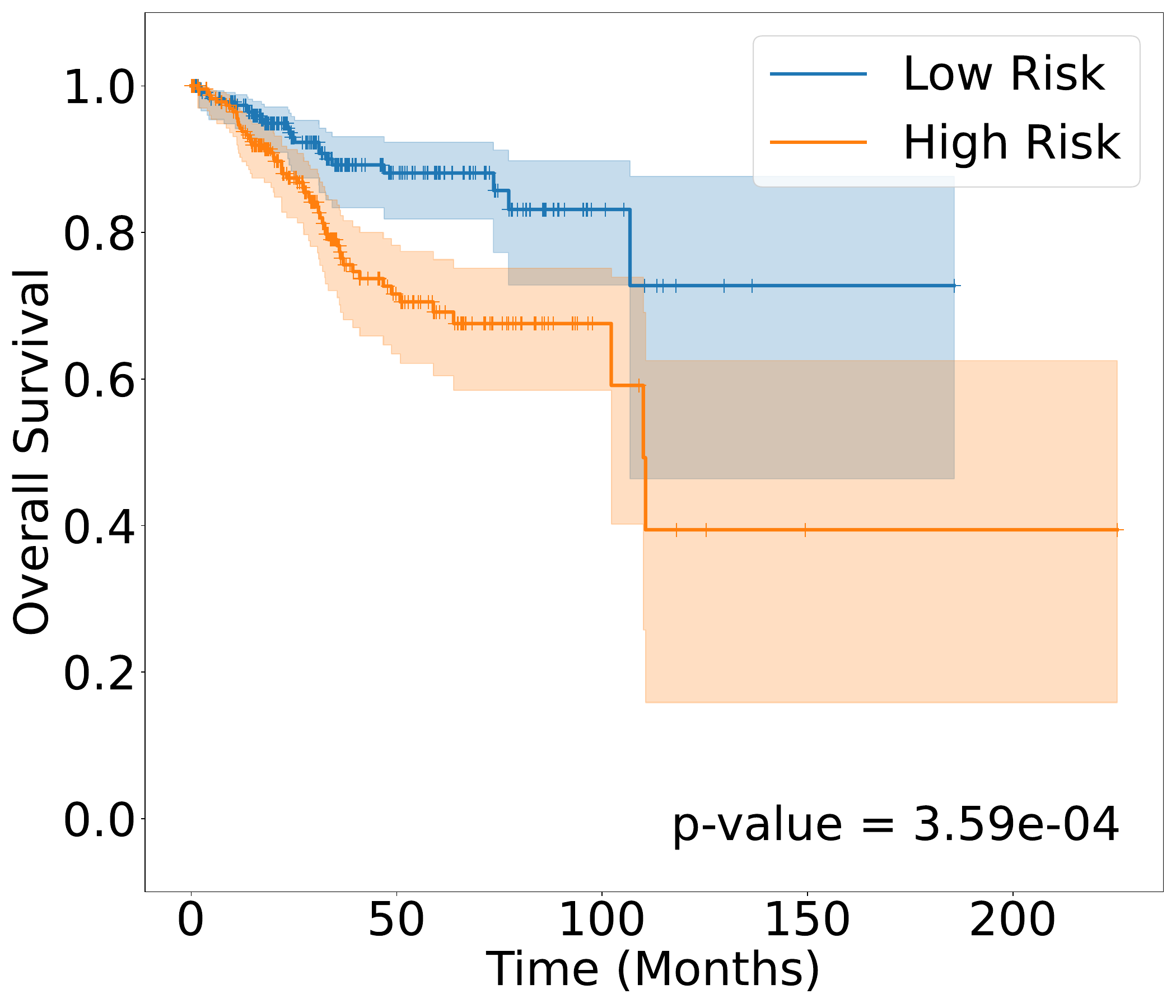}
\end{minipage}

\begin{minipage}{1 \textwidth}
    \begin{rotate}{90}
    \begin{minipage}{0.18 \textwidth} \centering \end{minipage}
    \end{rotate}
    \begin{minipage}{0.195 \textwidth} \centering \scriptsize (a) BLCA \end{minipage}
    \begin{minipage}{0.195 \textwidth} \centering \scriptsize (b) BRCA \end{minipage}
    \begin{minipage}{0.195 \textwidth} \centering \scriptsize (c) GBMLGG \end{minipage}
    \begin{minipage}{0.195 \textwidth} \centering \scriptsize (d) LUAD \end{minipage}
    \begin{minipage}{0.195 \textwidth} \centering \scriptsize (e) UCEC \end{minipage}
\end{minipage}

\caption{Kaplan-Meier curves of the proposed framework on five TCGA datasets.}
\label{fig:kmcurve}
\end{figure*}

\begin{figure*}[t]
\centering
\begin{minipage}{1 \textwidth}
    \includegraphics[width=0.195 \textwidth]{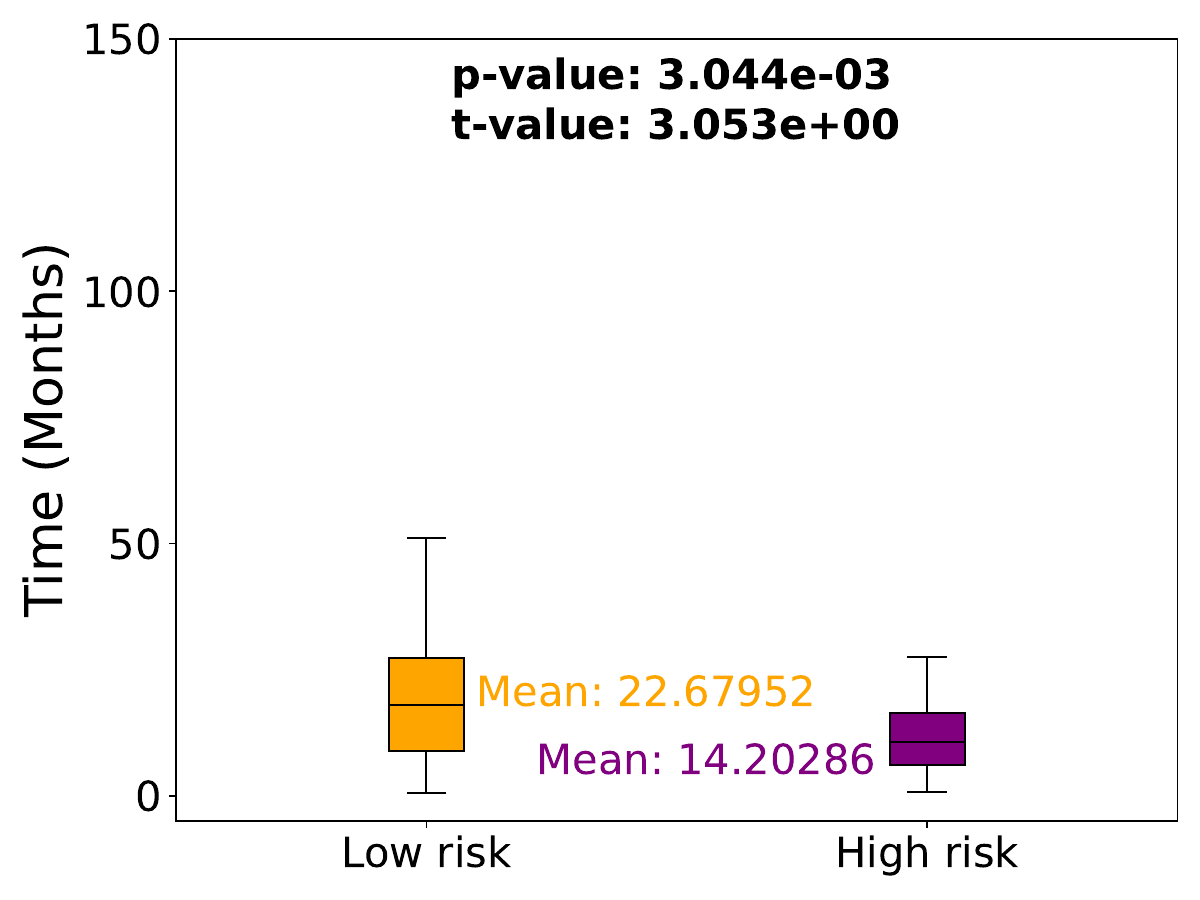}
    \includegraphics[width=0.195 \textwidth]{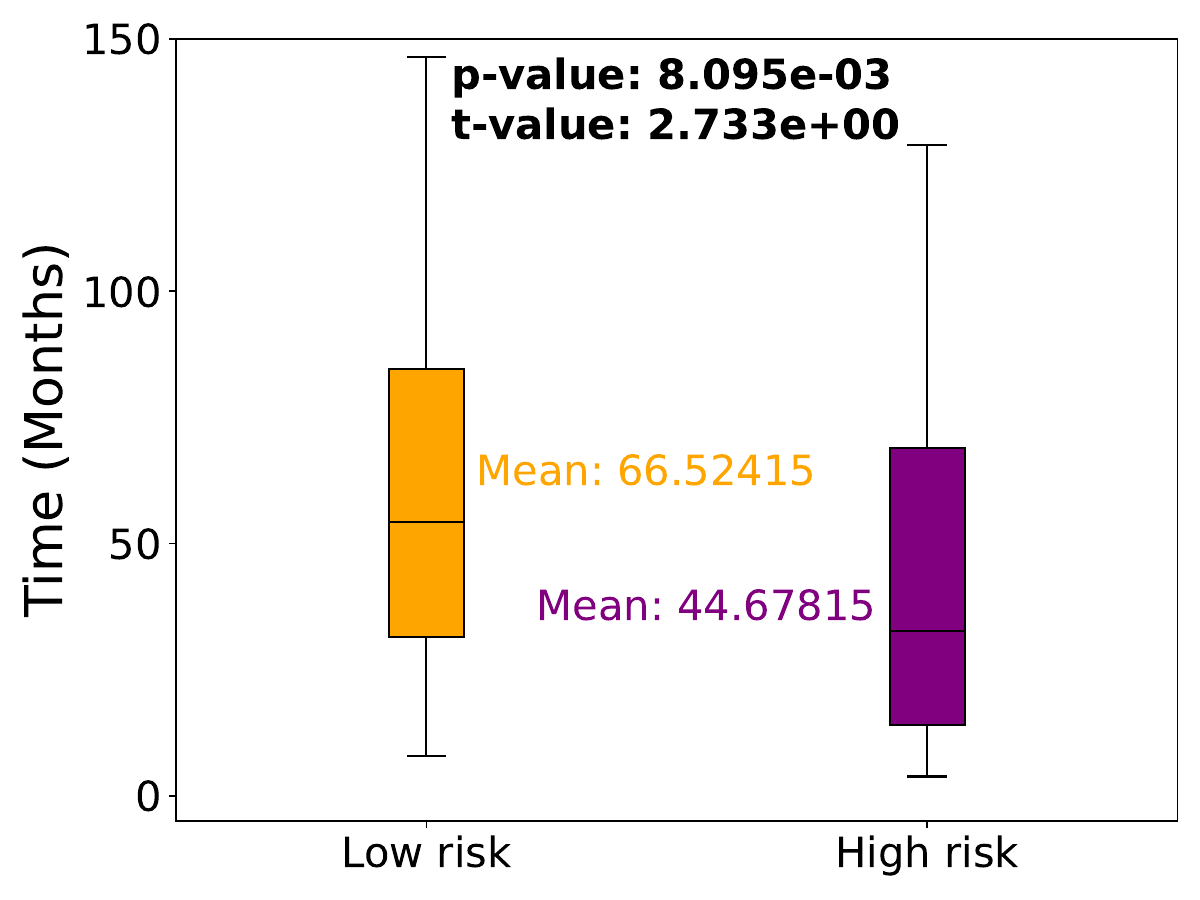}
    \includegraphics[width=0.195 \textwidth]{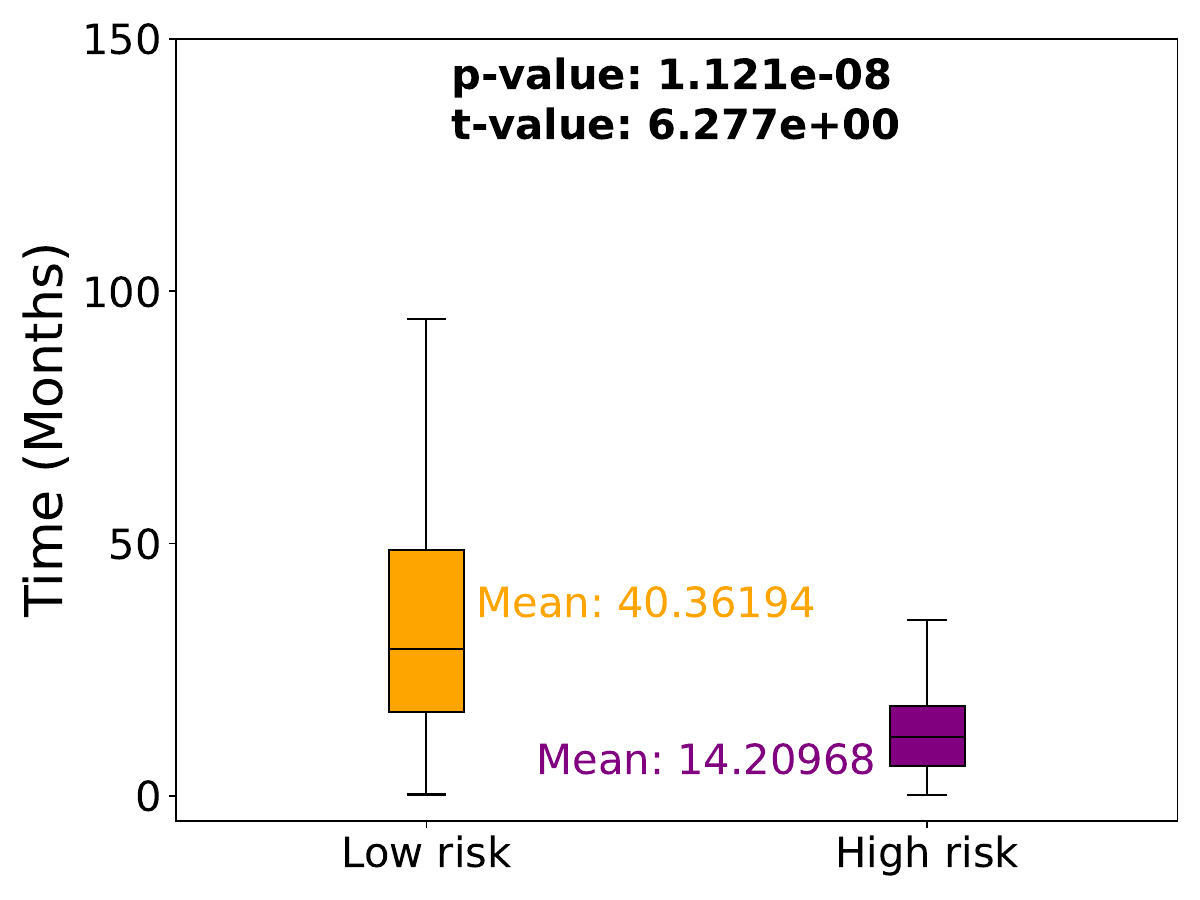}
    \includegraphics[width=0.195 \textwidth]{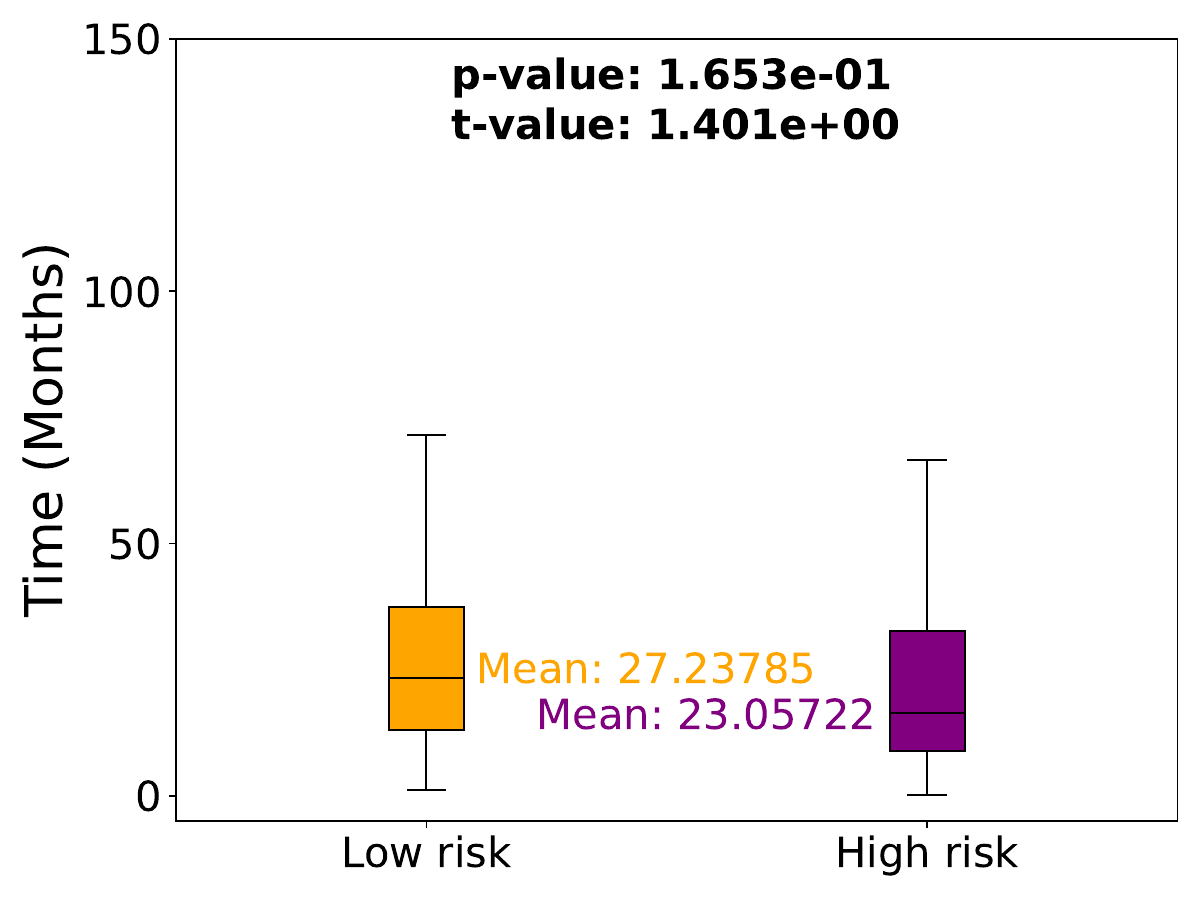}
    \includegraphics[width=0.195 \textwidth]{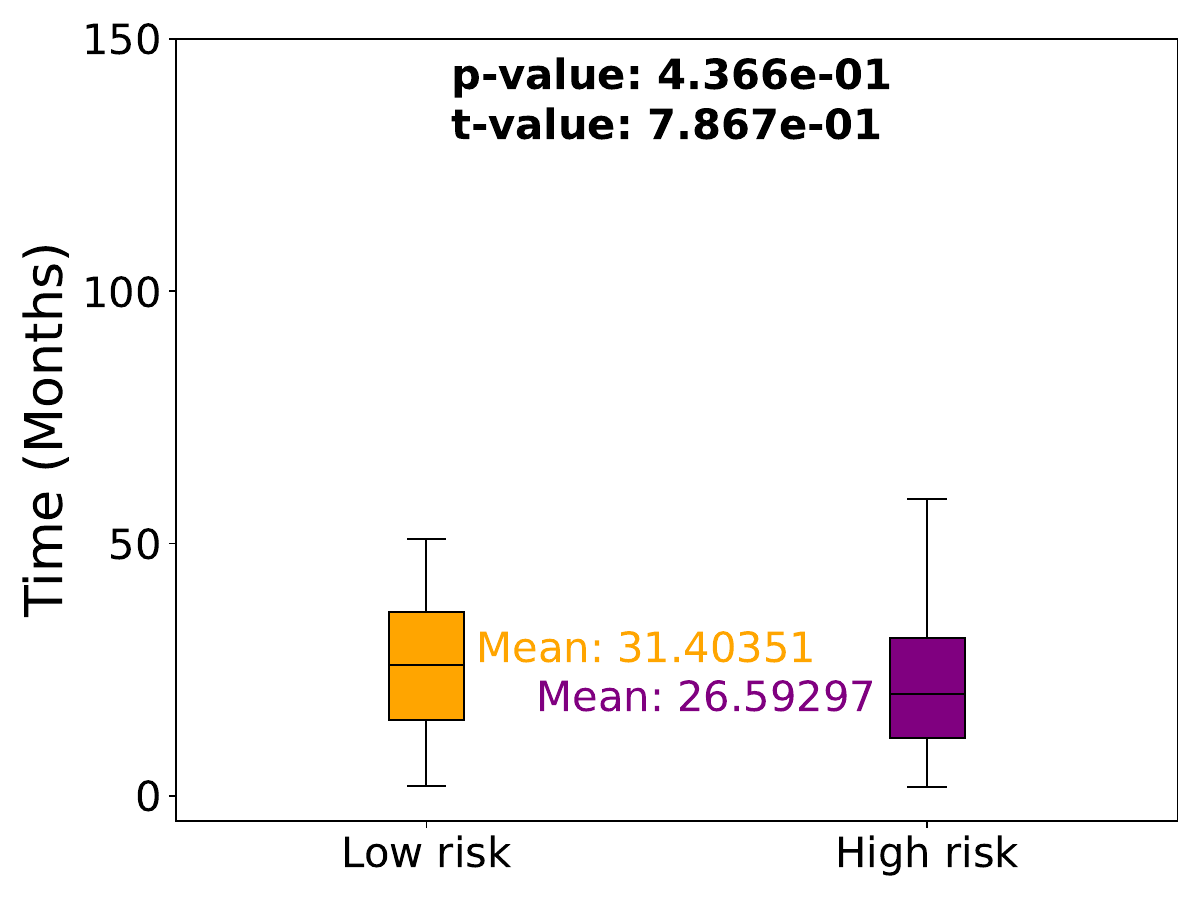}
\end{minipage}

\begin{minipage}{1 \textwidth}
    \begin{minipage}{0.195 \textwidth} \centering \scriptsize (a) BLCA \end{minipage}
    \begin{minipage}{0.195 \textwidth} \centering \scriptsize (b) BRCA \end{minipage}
    \begin{minipage}{0.195 \textwidth} \centering \scriptsize (c) GBMLGG \end{minipage}
    \begin{minipage}{0.195 \textwidth} \centering \scriptsize (d) LUAD \end{minipage}
    \begin{minipage}{0.195 \textwidth} \centering \scriptsize (e) UCEC \end{minipage}
\end{minipage}

\caption{T-test analysis of the proposed framework on five TCGA datasets.}
\label{fig:ttest}
\end{figure*}

\item \textbf{MOTCat} \cite{Xu_2023_ICCV} contains a global structure consistency, in which optimal transport (OT) is applied to match WSI patches and gene embeddings for selecting informative patches. More importantly, OT-based co-attention provides a global awareness to effectively capture structural interactions for survival prediction.
\item \textbf{CMTA} \cite{zhou2023cross} explores the intrinsic cross-modal correlations and transfers potential complementary information by enhancing modality-specific representations by integrating with multimodal representations. 
\end{itemize}
In addition, we build several models for a comprehensive comparison, including 1) SNNTrans: two SNN layers followed by a transformer; 2) MaxMIL/MeanMIL: a maximum/average operation to aggregate multiple instances; and 3) DualTrans: combining SNN and TransMIL for genomic profile and pathological images, respectively, while a Transformer is employed for multimodal fusion.

\subsection{Quantitative Evaluation}
\textbf{C-index Comparison.} 
As shown in Tab. \ref{tab:main}, the proposed framework achieves an average score of 72.60\%, which surpasses existing methods by a significant margin.
On each dataset, our model obtains 68.62\%, 68.40\%, 86.14\%, 69.57\%, and 70.26\% C-index scores on BLCA, BRCA, GBMLGG, LUAD, and UCEC, which are state-of-the-art performance on most datasets, while the second best on BLCA dataset with a small margin to the best 69.10\% reported by CMTA \cite{zhou2023cross}. 
Compared to unimodal models, either genomics- or pathology-based, our framework achieves significant improvements on all datasets.
It indicates that our framework can effectively integrate complementary information in genomic profiles and pathology images.
Compared to multimodal models, our framework reports a more generalized performance on survival analysis.
For example, CMTA \cite{zhou2023cross} performs well on the BLCA dataset but is still inferior to our method on others.

To prove the robustness of our framework, we compare the proposed framework with main competitors, including MCAT \cite{mcat}, SurvPath \cite{jaume2024modeling}, MOTCat \cite{Xu_2023_ICCV}, and CMTA \cite{zhou2023cross} by using latest pathology foundation models, including HIPT \cite{chen2022scaling}, PLIP \cite{huang2023visual}, UNI \cite{chen2024uni}, and CONCH \cite{lu2024visual}.
Our framework achieves the best overall performance when using HIPT, PLIP, and UNI as pathology feature extractors, respectively. 
Furthermore, our framework remains competitive to the best performance achieved by CMTA when using CONCH. 
These experiments further prove the robustness of the proposed framework.

\textbf{Kaplan-Meier Analysis.} 
The Kaplan-Meier (KM) analysis is a non-parametric statistic used to estimate the survival function from lifetime data.
Specifically, we use the median survival time of the entire cohort to divide all patients into high- and low-risk groups, represented as red and blue curves, respectively.
The p-value in KM analysis indicates the probability of observing the observed difference in survival rates between two groups, under the null hypothesis that there is no true difference between them.
As shown in Fig. \ref{fig:kmcurve}, the p-values achieved by our framework are significantly lower than 0.05 on all five datasets, which indicates a statistically significant discrimination between high- and low-risk groups.
We also compare the p-values with several representative methods in Tab. \ref{tab:pvalue}.
Our model achieves the lowest p-values on BLCA and BRCA and, meanwhile is competitive to state-of-the-art methods on GBMLGG, LUAD, and UCEC.
Our framework can generalize across different cancers to produce robust predictions, holding the potential to significantly advance both clinical practice and cancer research.

\begin{table}[!t]
\centering
\caption{Comparison between p-values of Kaplan-Meier analysis.}
\setlength\tabcolsep{5pt}
\label{tab:pvalue}
\begin{tabular}{c|ccccc}
        \toprule
Methods  & BLCA   & BRCA   & GBMLGG & LUAD   & UCEC    \\\midrule
SNN      & $1.7e^{-6}$ & $1.5e^{-2}$ & $1.1e^{-27}$ & $1.1e^{-3}$ & $9.8e^{-4}$  \\
TransMIL & $5.6e^{-2}$ & $2.4e^{-3}$ & $2.5e^{-25}$ & $3.8e^{-2}$ & $9.8e^{-2}$  \\
MCAT     & $7.8e^{-6}$ & $7.5e^{-3}$ & $1.6e^{-19}$ & $6.9e^{-5}$ & $4.8e^{-3}$  \\
SurvPath & $8.2e^{-6}$ & $1.4e^{-3}$ & $1.0e^{-29}$ & $2.2e^{-4}$ & $1.4e^{-5}$  \\
MOTCat   & $2.9e^{-7}$  & $4.9e^{-4}$  & $3.4e^{-30}$ & $1.1e^{-5}$ & $\mathbf{3.8e^{-7}}$  \\
CMTA     & $2.0e^{-8}$  & $3.1e^{-3}$  & $\mathbf{1.8e^{-33}}$ & $\mathbf{9.6e^{-7}}$ & $1.7e^{-3}$  \\
Ours     & $\mathbf{5.7e^{-11}}$ & $\mathbf{2.2e^{-7}}$  & $9.8e^{-32}$ & $1.1e^{-4}$ & $3.6e^{-4}$  \\\bottomrule
\end{tabular}
\end{table}

\textbf{T-test Analysis.} 
The T-test is a statistical analysis method used to determine if there is a significant difference between the means of two groups.
In the T-test analysis, the t-value quantifies the size of the difference relative to the variability within two groups, while the p-value is the probability that the results from our results occurred by chance.  
As the results shown in Fig. \ref{fig:ttest}, our framework can distinguish between different groups with good p-values and t-values, especially on BLCA, BRCA, and GBMLGG datasets (p-value \textless \: 0.05).
However, LUAD and UCEC datasets are still challenging for our framework (p-value \textgreater \: 0.05). 

\textbf{Efficiency Comparison.} 
Since all compared survival models require extracting pathology patch embeddings beforehand, we exclude the embedding extraction process in our efficiency comparison in Tab. \ref{tab:resource}.
Our framework demands increased GPU resources and computation time compared to previous methods, primarily due to the inclusion of online K-means clustering. 
However, by employing K-means clustering in advance, the GPU memory consumption of our pure network (denoted with *) is considerably lower than existing methods, while achieving faster processing speeds than state-of-the-art CMTA and MOTCat methods.

\begin{table}[!t]
\centering
\caption{Computational resource comparison. Training and test time measures the numbers of milliseconds for processing a single sample. * means that K-means is employed to cluster pathology patches beforehand.}
\setlength\tabcolsep{10pt}
   \label{tab:resource}
   \begin{center}
      \begin{tabular}{c|ccccc|c}
        \toprule
         Methods  &   GPU memory   & Training time  & Test time    \\
        \midrule
         MCAT \cite{mcat}                    & 4.06  & 13.7  & 9.6 \\
         SurvPath \cite{jaume2024modeling}   & 1.95  & \textbf{8.9}   & \textbf{5.3} \\
         MOTCat \cite{Xu_2023_ICCV}          & 3.09  & 39.2  & 38.3 \\
         CMTA  \cite{zhou2023cross}          & 19.70 & 33.1  & 28.1 \\
         Ours*                               & \textbf{0.44}  & 21.6  & 16.1 \\
         Ours                                & 18.9  & 95.2  & 81.1 \\
        \bottomrule
      \end{tabular}
   \end{center}
\end{table}

\begin{table}[!t]
\centering
\caption{Ablation study on the proposed modules. CCA, MKD and CGM indicate the proposed cluster center alignment, multimodal knowledge decomposition and cohort guidance modeling, respectively.}
\setlength\tabcolsep{2pt}
\label{tab:component}
\begin{tabular}{ccc|ccccc|c}
\toprule
CCA & MKD & CGM                & BLCA   & BRCA   & GBMLGG & LUAD   & UCEC   & Overall  \\\midrule
\multicolumn{3}{c|}{(Baseline)}        & 0.6607 & 0.6637  & 0.8393  & 0.6706  & 0.6724  & 0.7013 \\
\checkmark &&                          & 0.6701 & 0.6767  & 0.8433  & 0.6792  & 0.6844  & 0.7107 \\
\checkmark & \checkmark &              & 0.6783 & 0.6718  & 0.8511  & 0.6855  & 0.6872  & 0.7147 \\\hline
\checkmark & \checkmark &  \checkmark  & \textbf{0.6862} & \textbf{0.6840}  & \textbf{0.8614}  & \textbf{0.6957}  & \textbf{0.7026}  & \textbf{0.7260} \\\bottomrule
\end{tabular}
\end{table}

\subsection{Ablation study}
\textbf{Module Analysis.} 
To demonstrate the effectiveness of our methods, we progressively append our modules to our baseline DualTrans, including Cluster Center Alignment (CCA), Multimodal Knowledge Discrimination (MKD), and Cohort Guidance Modeling (CGM).
As shown in Tab. \ref{tab:component}, the proposed methods effectively and consistently improve the performance of the baseline DualTrans model.
CCA aligns the cluster centers after K-means, facilitating the extraction of crucial information from important patches.
Moreover, MKD enables more effective multimodal fusion by comprehensively decomposing multimodal knowledge, reducing redundancy, and reinforcing valuable information.
Further, CGM leverages the cohort guidance to promote general multimodal interaction learning. 
Collaborating with these modules, our full framework achieves significant improvements over the baseline model.

\begin{figure*}[!t]
\centering
\includegraphics[width=0.325 \textwidth]{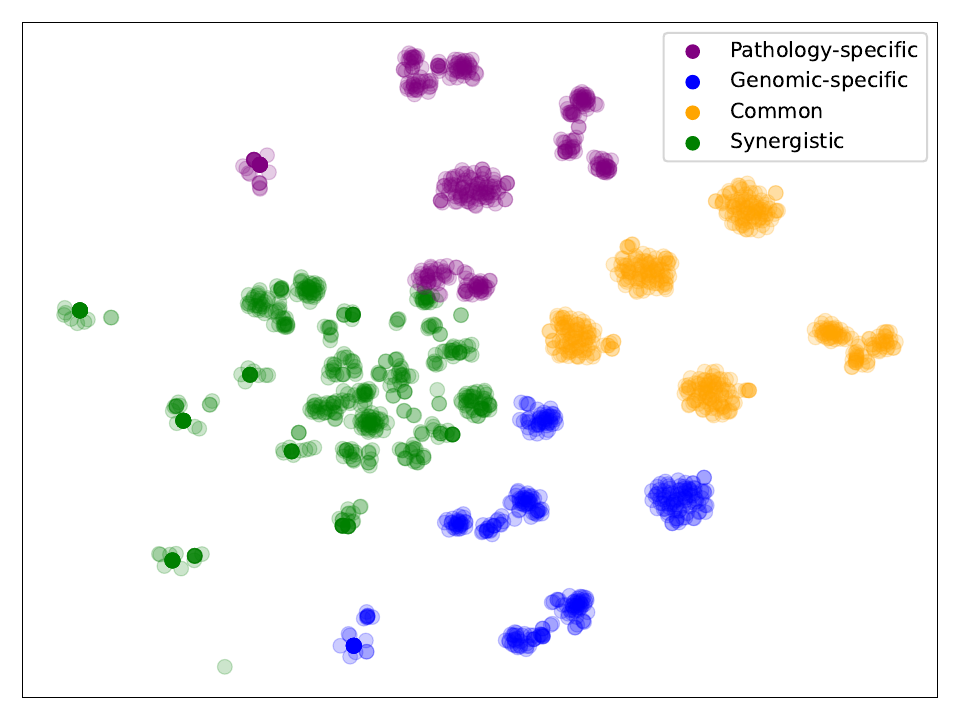}
\includegraphics[width=0.325 \textwidth]{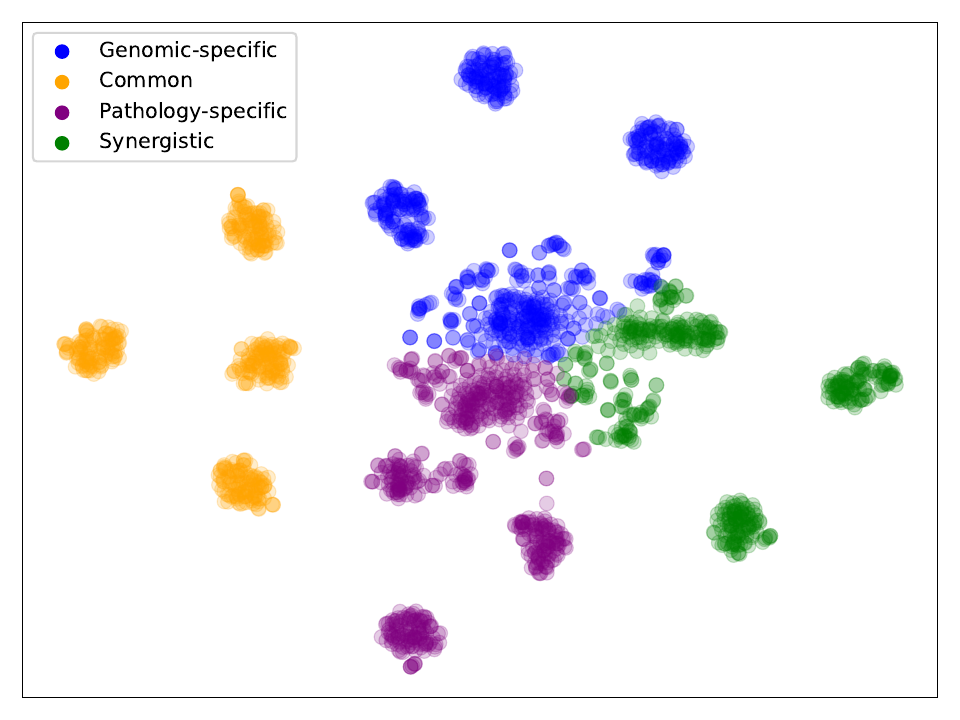}
\includegraphics[width=0.325 \textwidth]{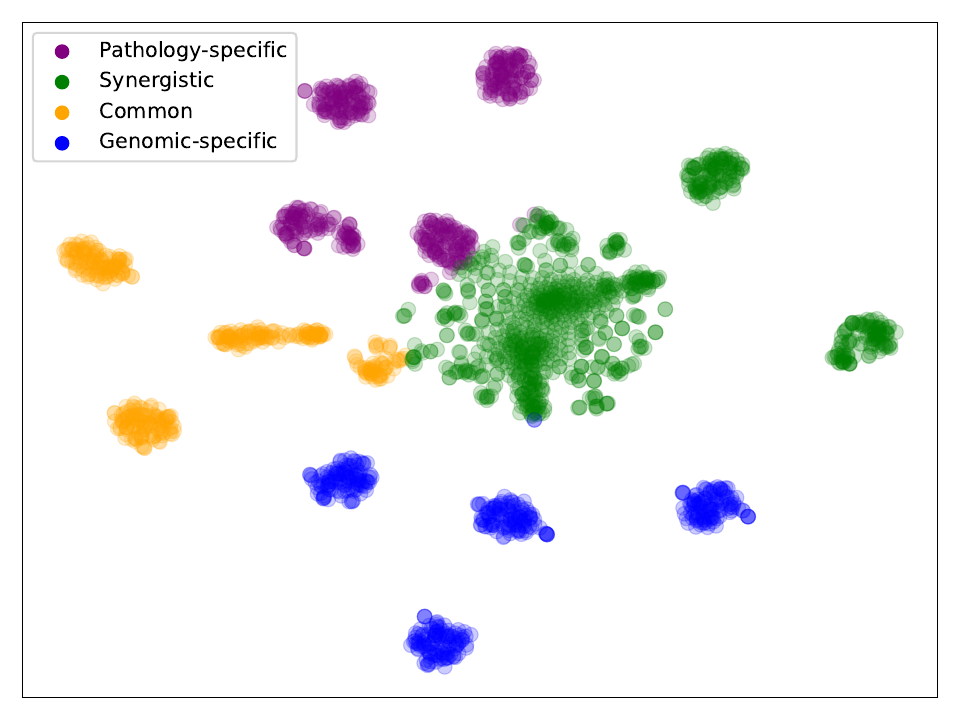}
\begin{minipage}{0.325 \textwidth} \centering  (a) BLCA \end{minipage}
\begin{minipage}{0.325 \textwidth} \centering  (b) BRCA \end{minipage}
\begin{minipage}{0.325 \textwidth} \centering  (c) GBMLGG \end{minipage}
\begin{minipage}{0.99 \textwidth}
\centering
\includegraphics[width=0.325 \textwidth]{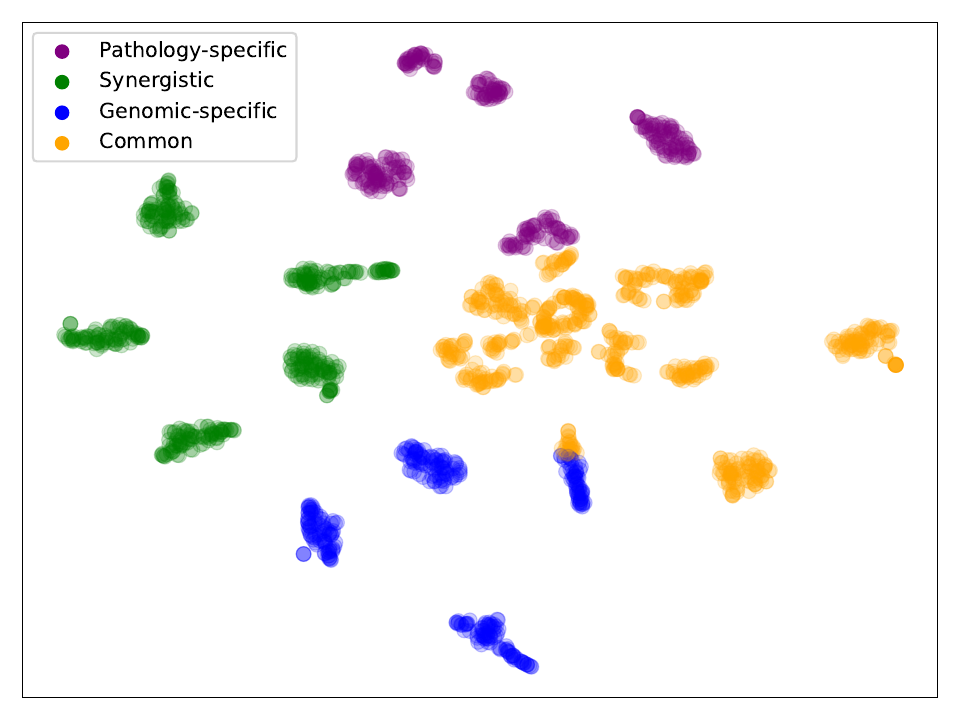}
\includegraphics[width=0.325 \textwidth]{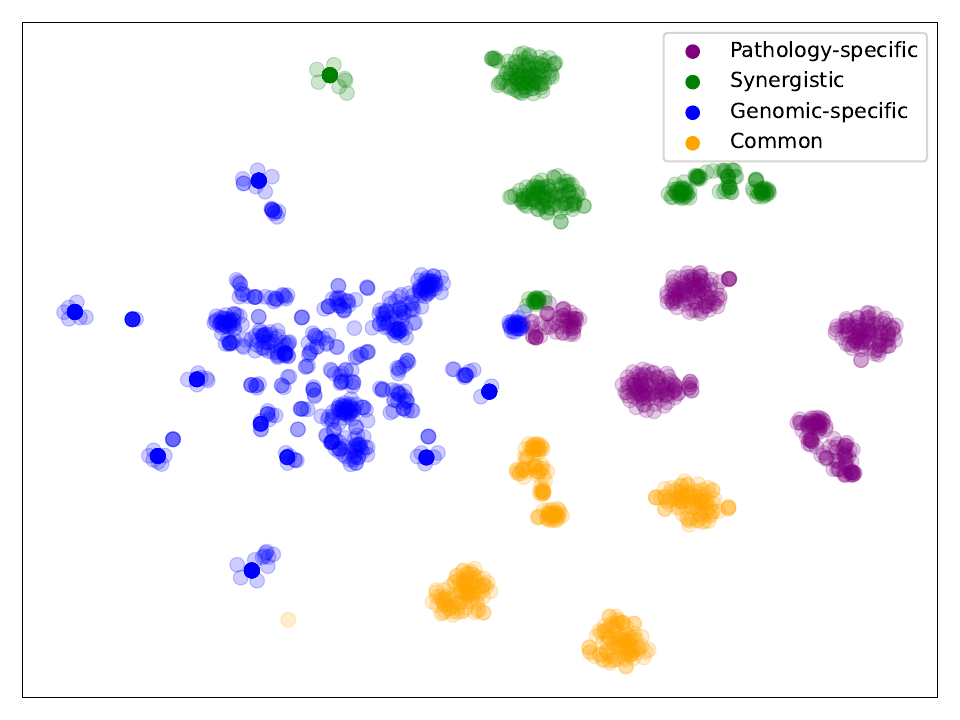}
\end{minipage}
\begin{minipage}{0.325 \textwidth} \centering  (d) LUAD \end{minipage}
\begin{minipage}{0.325 \textwidth} \centering  (e) UCEC \end{minipage}
\caption{T-SNE visualization of the knowledge distributions.}
\label{fig:knowledge}
\end{figure*}

\textbf{Impact of Cluster Alignment.}
To prove the necessity of cluster alignment, we calculate the mean L2 distances among cluster representations at the same ordinal position before and after our alignment process in Tab. \ref{tab:dist}.
Our alignment method significantly reduces the distances between cluster representations. 
However, there still exist considerable distances between these representations, mainly due to the different patch distributions and personalized information of patients.
Addressing this issue may necessitate the development of more sophisticated approaches in future research.

\textbf{Impact of Knowledge Encoders.}
In our framework, we employ two encoders to learn common and synergistic knowledge, respectively.
As introduced in previous sections and many related works \cite{williams2010nonnegative, liang2023foundations, liang2024quantifying, liang2023multimodal}, the knowledge in multimodal fusion may be one of common, synergistic, and two modality-specific components, serving as the theoretical foundation of our framework structure.
If more co-attention encoders are introduced, they will not fall out of these four knowledge components, becoming redundant information.
To prove this point, we provide the results of our framework with more co-attention encoders in Tab. \ref{tab:encoder}, where two of them are supervised as common and synergistic encoders and others have no specific supervision signals.
Furthermore, we also validate our framework with removed synergistic or common encoders.
With more encoders employed, we have not seen substantial improvements in overall performance.
Meanwhile, since the integrity of multimodal knowledge is destroyed, we witness varying performance degradation on different datasets.

\begin{table}[!t]
\centering
\caption{Average distances of cluster representations before and after alignment.}
\setlength\tabcolsep{5pt}
   \label{tab:dist}
   \begin{center}
      \begin{tabular}{c|ccccc}
\toprule
         Distances  &   BLCA   & BRCA  & GBMLGG & LUAD & UCEC    \\
\midrule
         Before alignment & 4.6501 & 5.5141 & 7.1205 & 5.6450 & 4.6632 \\
         After alignment  & \textbf{3.4775} & \textbf{4.6234} & \textbf{6.0795} & \textbf{4.9767} & \textbf{3.8678} \\
\bottomrule
      \end{tabular}
   \end{center}
\end{table}

\begin{table}[!t]
\centering
\caption{Ablation study on our framework with different numbers of knowledge encoders.}
\setlength\tabcolsep{3pt}
   \label{tab:encoder}
   \begin{center}
      \begin{tabular}{c|ccccc|c}
\toprule
         \# of encoders &   BLCA   & BRCA  & GBMLGG & LUAD  & UCEC       &   Overall   \\
\midrule
          5 (Ours + 3)                                    & 0.6689               & \textbf{0.6931}               & 0.8513               & 0.6856               & \textbf{0.7089}         & 0.7216       \\
          3 (Ours + 1)                                   & \textbf{0.6883}               & 0.6798               & 0.8581               & \textbf{0.6967}              & 0.6998         & 0.7245       \\
          1 (Common)          & 0.6793  & 0.6851  & 0.8549        & 0.6772      & 0.6967     & 0.7186   \\
          1 (Synergistic)                                     & 0.6764               & 0.6489              & 0.8576             & 0.6752             & 0.6794         & 0.7075       \\
            2 (Ours)    & 0.6862  & 0.6840  & \textbf{0.8614}  & 0.6957  & 0.7026   & \textbf{0.7260} \\
\bottomrule
      \end{tabular}
   \end{center}
\end{table}

\textbf{Knowledge Component Visualization.} 
In Fig. \ref{fig:knowledge}, we visualize the feature distributions after MKD using T-SNE \cite{hinton2002stochastic}, where the orange, green, blue, and purple points indicate common, synergistic, genomic-specific, and pathology-specific knowledge, respectively. 
The distributions of these components differ from each other and are consistent with our intuition that common and synergistic features lie in the middle of two modality-specific features.
Moreover, the features of each component are clustered as several groups, which demonstrates that patients of similar risk tend to be closer than other patients under our cohort guidance.
For example, genomic-specific features are clustered into four groups, which is consistent with the group number used in our framework. 
It proves that our framework can extract discriminative representations to facilitate multimodal fusion. 

\textbf{Impact of $l_k$.}
An ablation study on the similarity constraints in $l\_k$ is conducted by removing the supervision signals for different encoders, respectively, as shown in Tab. \ref{tab:sim}.
Since different tasks may benefit more from different knowledge components, removing the similarity constraints causes performance degradation to different extents. 
It proves that, without explicit modeling of knowledge components, important knowledge in multimodal data is easy to be overlooked.

\begin{table}[!t]
\centering
\caption{Ablation study on our framework with different similarity constraints in $l_k$. G, P, C, and S indicate the constraints for corresponding knowledge components.}
\setlength\tabcolsep{2pt}
   \label{tab:sim}
   \begin{center}
      \begin{tabular}{c|ccccc|c}
\toprule
         Similarity constraints &   BLCA   & BRCA  & GBMLGG & LUAD  & UCEC       &  Overall    \\
        \midrule
        G+P+S               & 0.6598  & 0.6931  & 0.8647        & 0.6722      & 0.6793     & 0.7138       \\
        G+P+C               & 0.6784  & 0.6872  & \textbf{0.8657}        & 0.6868      & 0.6983     & 0.7233       \\
        P+C+S               & 0.6720  & \textbf{0.6970}  & 0.8489        & \textbf{0.7003}      & 0.6781     & 0.7193       \\
        G+C+S               & \textbf{0.6915}  & 0.6786  & 0.8586        & 0.6834      & \textbf{0.7057}     & 0.7236      \\
        G+P+C+S (Ours)      & 0.6862  & 0.6840  & 0.8614        & 0.6957      & 0.7026     & \textbf{0.7260} \\
        \bottomrule
      \end{tabular}
   \end{center}
\end{table}

\textbf{Hyper-parameter Experiment.} 
There are several hyper-parameters used in our framework, including $k$ in K-means, anchor updating ratio $\tau$, cohort bank length $b$, and the number of patient groups $r$.
We provide experiment results on these hyper-parameters in Tab. \ref{tab:hyper}.
From these results, we conclude that $k=6$, $\tau=0.1$, $b=10$, and $r=4$ work best in our framework, and thus are employed as the main settings of our framework.
Moreover, smaller values of $k$, $b$, and $r$ imply fewer features and computational loads, but the performance usually drops significantly.
When exceeding adequate values, enlarging them may not offer additional improvements.
For $\tau$, it controls the updating rate of the anchor, which is used in cluster center alignment.
After a period of training, the anchor becomes stable, so the effect of $\tau$ may not be significant.

\begin{table}[!t]
\centering
\caption{Results of hyper-parameter experiments.}
\setlength\tabcolsep{3pt}
\label{tab:hyper}
\begin{tabular}{c|c|ccccc|c}
\toprule
Param. & Value & BLCA   & BRCA   & GBMLGG & LUAD   & UCEC & Overall \\\midrule
    & 4 & 0.6758 & 0.6763 & 0.8433 & 0.6718 & 0.6775  & 0.7089 \\
$k$ & 6 & \textbf{0.6862} & 0.6840 & \textbf{0.8614} & 0.6957 & \textbf{0.7026}  & \textbf{0.7260} \\
    & 9 & 0.6792 & \textbf{0.6851} & 0.8573 & \textbf{0.6960} & 0.6982  & 0.7232 \\\hline
       & 0.05 & 0.6834 & 0.6842 & 0.8598 & \textbf{0.6957} & 0.6997  & 0.7246 \\
$\tau$ & 0.1  & \textbf{0.6862} & 0.6840 & \textbf{0.8614} & \textbf{0.6957} & \textbf{0.7026}  & \textbf{0.7260} \\
       & 0.3  & 0.6855 & \textbf{0.6847} & 0.8602 & 0.6932 & 0.7011  & 0.7249 \\\hline
    & 5  & 0.6764 & 0.6799 & \textbf{0.8631} & 0.6789 & 0.6911  & 0.7179 \\
$b$ & 10 & \textbf{0.6862} & 0.6840 & 0.8614 & \textbf{0.6957} & \textbf{0.7026}  & \textbf{0.7260} \\
    & 20 & 0.6835 & \textbf{0.6844} & 0.8597 & 0.6827 & 0.6943  & 0.7209 \\\hline
    & 2 & 0.6778 & 0.6751 & 0.8501 & 0.6747 & 0.6818  & 0.7119 \\
$r$ & 4 & \textbf{0.6862} & 0.6840 & \textbf{0.8614} & \textbf{0.6957} & \textbf{0.7026}  & \textbf{0.7260} \\
    & 6 & 0.6754 & \textbf{0.6912} & 0.8486 & 0.6913 & 0.6988  & 0.7211 \\\bottomrule
\end{tabular}
\end{table}

\begin{figure*}[!t]
\centering
\begin{minipage}{1 \textwidth}
    \includegraphics[width=0.25 \textwidth]{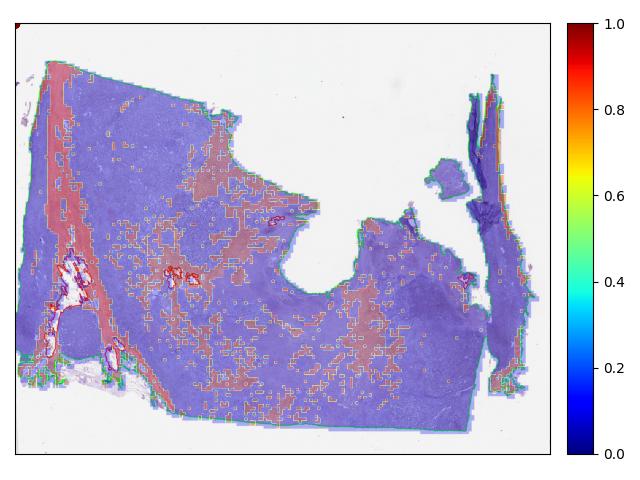}
    \includegraphics[width=0.25 \textwidth]{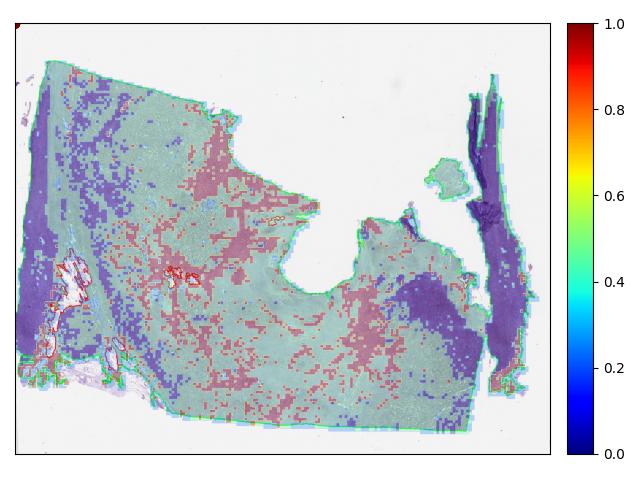}
    \includegraphics[width=0.25 \textwidth]{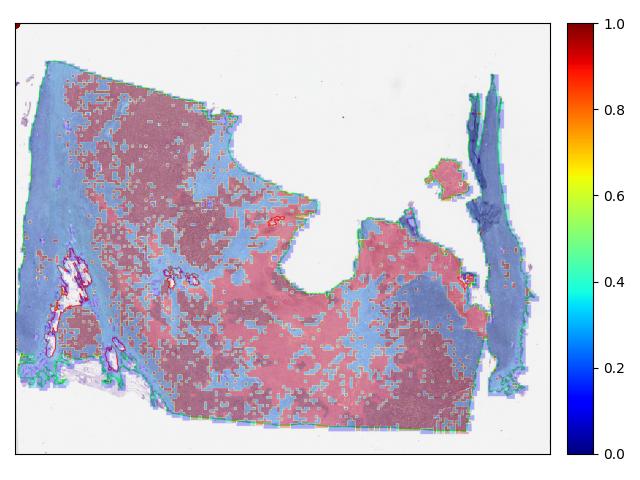}
    \includegraphics[width=0.23 \textwidth]{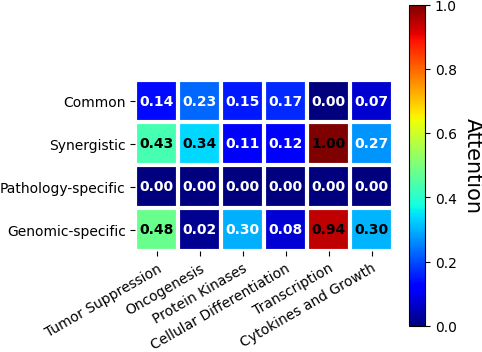}
\end{minipage}
\begin{minipage}{1 \textwidth}
\vspace{0.1in}
    \includegraphics[width=0.25 \textwidth]{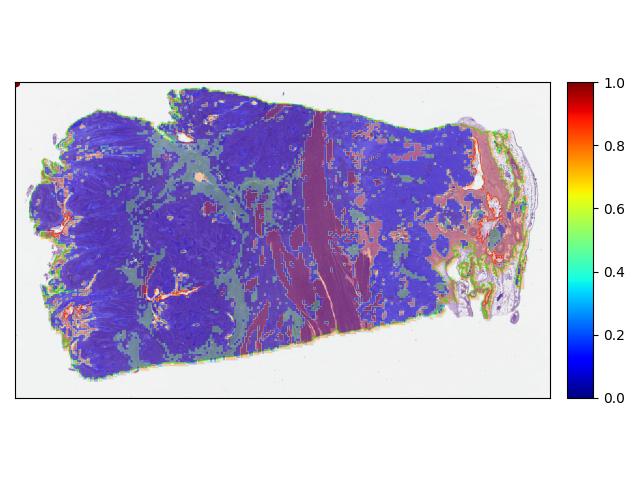}
    \includegraphics[width=0.25 \textwidth]{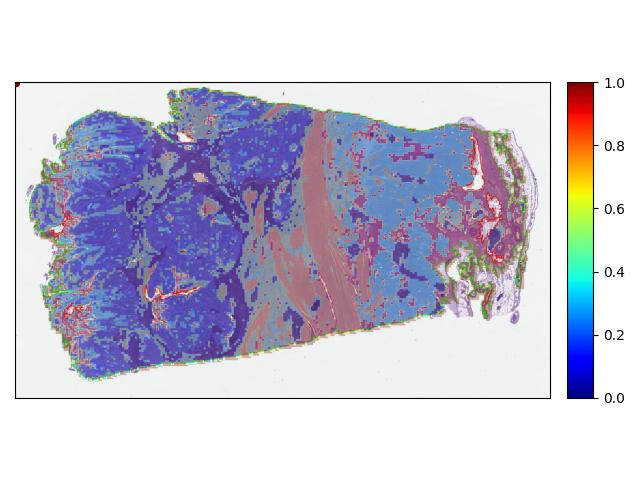}
    \includegraphics[width=0.25 \textwidth]{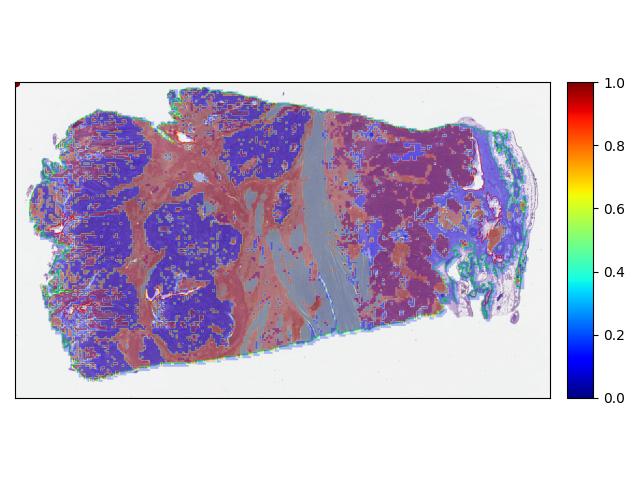}
    \includegraphics[width=0.23 \textwidth]{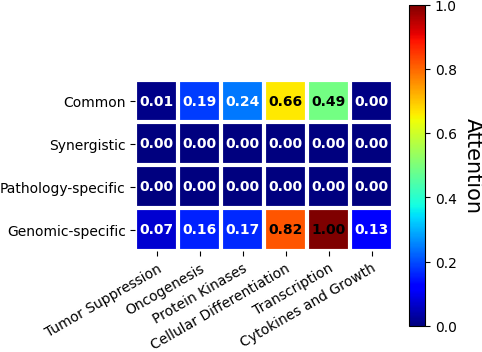}
\end{minipage}
\begin{minipage}{1 \textwidth}
\vspace{0.1in}
    \begin{minipage}{0.25 \textwidth} \centering \scriptsize (a) Common knowledge \end{minipage}
    \begin{minipage}{0.25 \textwidth} \centering \scriptsize (b) Synergistic knowledge \end{minipage}
    \begin{minipage}{0.25 \textwidth} \centering \scriptsize (c) Pathology-specific knowledge \end{minipage}
    \begin{minipage}{0.23 \textwidth} \centering \scriptsize (d) Attention on genomics \end{minipage}
\end{minipage}

\caption{Attention visualization of different knowledge components on input modalities.}
\label{fig:visual}
\end{figure*}

\textbf{Knowledge Attention Analysis.} 
To further illustrate the differences between knowledge components, we visualize their attention on input data in Fig. \ref{fig:visual}.
In sub-figures (a)-(c), we visualize the attention regions of redundancy, synergy, and pathology-specific knowledge on input pathology images, while the last sub-figure (d) shows attention scores on genomic profiles.
For pathology images, the focusing regions vary from component to component, indicating the distinctiveness between them.
For genomic profiles, the first sample focuses more on synergy, whereas the second one has a stronger reliance on redundancy.
It implies that two types of multimodal interactions are necessary for effective survival analysis.

\section{Conclusion}
In this paper, we propose a Cohort-individual Cooperative Learning (CCL) framework to effectively integrate genomics and pathology images for cancer survival analysis.
Specifically, first, we propose a Multimodal Knowledge Decomposition (MKD) module to completely decompose multimodal knowledge into four distinct components. 
Second, we propose a Cohort Guidance Modeling (CGM) to enhance the generalization and discrimination abilities of the decomposed components.
Collaborating knowledge decomposition and cohort guidance, our framework can learn general multimodal interactions and facilitate an effective fusion.
Experiment results on five TCGA datasets prove that the proposed framework achieves state-of-the-art performance in survival analysis.

Despite the performance achieved by our framework, there are several potential directions for future work.
First, a more general cluster center alignment approach can be designed to handle patients with significantly different patch distributions.
Second, soft similarity assessment among patients may improve the cohort guidance by providing a more effective learning strategy.
Third, specialized encoders for different knowledge components may further facilitate the multimodal fusion process. 
Last but not least, more modalities describing other characteristics of tumors, such as diagnostic reports and clinical indicators, can be utilized in survival analysis models.

\bibliographystyle{IEEEtran}
\bibliography{main}

\end{document}